\documentclass[lettersize,journal]{IEEEtran}
\usepackage{amsmath,amsfonts}
\usepackage{algorithmic}
\usepackage{algorithm}
\usepackage{array}
\usepackage{textcomp}
\usepackage{stfloats}
\usepackage{url}
\usepackage{verbatim}
\usepackage{graphicx}
\usepackage{cite}
\usepackage{multirow}
\usepackage{subfigure}
\usepackage{bm}
\usepackage{xcolor}
\usepackage{makecell}
\usepackage{booktabs} 
\usepackage{siunitx}
 \usepackage{amssymb}
 \usepackage{fancyhdr}
  \usepackage{hyperref}
\fancypagestyle{mainFancy}{
    \fancyhf{}

    \fancyhead[C]{\fzkai{\leftmark}}       
    \fancyhead[C]{\sffamily \fontsize{8}{10}\selectfont This article has been accepted for publication in IEEE/ACM Transactions on Audio, Speech and Language Processing. This is the author's version which has not been fully edited and
content may change prior to final publication. Citation information: DOI 10.1109/TASLP.2023.3277693}
    \fancyfoot[C]{\sffamily \fontsize{8}{10}\selectfont© 2023 IEEE. Personal use is permitted, but republication/redistribution requires IEEE permission.} 
}
\begin{document}

\title{Latent-Domain Predictive Neural Speech Coding}

\author{Xue Jiang, Xiulian Peng, Huaying Xue, Yuan Zhang, Yan Lu
\thanks{Xue Jiang is with the School of Information and Communication Engineering, Communication University of China, Beijing 100024, China (e-mail: jiangxhoho@cuc.edu.cn).}
\thanks{X. Peng, H. Xue and Y. Lu are with the Microsoft Research Asia, Beijing 100080, China (e-mail: xipe@microsoft.com; huxue@microsoft.com; yanlu@microsoft.com).}
\thanks{Y. Zhang is with the State Key Laboratory of Media Convergence and Communication, Communication University of China, Beijing 100024, China (e-mail: yzhang@cuc.edu.cn).}
\thanks{Yan Lu and Xiulian Peng are corresponding authors.}
\thanks{This work was done when Xue Jiang was an intern at Microsoft Research Asia.}}



\maketitle\thispagestyle{fancy}

\begin{abstract}
Neural audio/speech coding has recently demonstrated its capability to deliver high quality at much lower bitrates than traditional methods. However, existing neural audio/speech codecs employ either acoustic features or learned blind features with a convolutional neural network for encoding, by which there are still temporal redundancies within encoded features. This paper introduces latent-domain predictive coding into the VQ-VAE framework to fully remove such redundancies and proposes the TF-Codec for low-latency neural speech coding in an end-to-end manner. Specifically, the extracted features are encoded conditioned on a prediction from past quantized latent frames so that temporal correlations are further removed. Moreover, we introduce a learnable compression on the time-frequency input to adaptively adjust the attention paid to main frequencies and details at different bitrates. A differentiable vector quantization scheme based on distance-to-soft mapping and Gumbel-Softmax is proposed to better model the latent distributions with rate constraint. Subjective results on multilingual speech datasets show that, with low latency, the proposed TF-Codec at 1 kbps achieves significantly better quality than Opus at 9 kbps, and TF-Codec at 3 kbps outperforms both EVS at 9.6 kbps and Opus at 12 kbps. Numerous studies are conducted to demonstrate the effectiveness of these techniques. Code and models are available at \href{https://github.com/microsoft/TF-Codec}{https://github.com/microsoft/TF-Codec}.
\end{abstract}


\section{Introduction}
In recent years, neural audio/speech coding has rapidly advanced and now delivers high-quality results at very low bitrates, particularly for speech. Existing neural codecs {can} mainly be divided into two categories, generative decoder model-based codecs \cite{wavcodec,Lyra,sampleRNNcodec,generative,improveopus} and end-to-end neural audio/speech coding \cite{VQ-VAE-wavenet,disentangle,soundstream,cascaded,scalable-codec,Configurablecodec,TFNetCodec}. The former extracts acoustic features from the audio for encoding and employs a powerful decoder to reconstruct the waveform based on generative models. The latter mainly leverages the VQ-VAE \cite{VQ-VAE} framework to learn an encoder, a vector quantizer and a decoder in an end-to-end manner. The latent features to be quantized are mostly blindly learned using a convolutional neural network (CNN) without any prior knowledge. These methods have largely improved the coding efficiency by achieving high quality at low bitrates. However, temporal correlations are not fully exploited in these algorithms, resulting in much redundancy among neighboring frames in encoded features. In light of this, we propose incorporating predictive coding into the VQ-VAE-based neural coding framework to remove such redundancies. 

\begin{figure}[tb]
\begin{minipage}[b]{1.0\linewidth}
  \centering
  \centerline{\includegraphics[width=8.7cm]{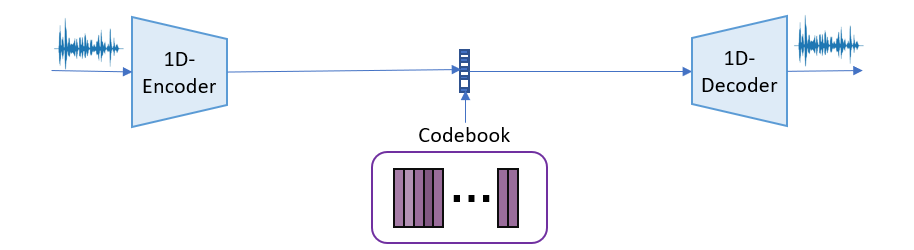}}
  \centerline{(a) Previous VQ-VAE based neural codec}\medskip
\end{minipage}
\begin{minipage}[b]{1.0\linewidth}
  \centering
  \centerline{\includegraphics[width=8.5cm]{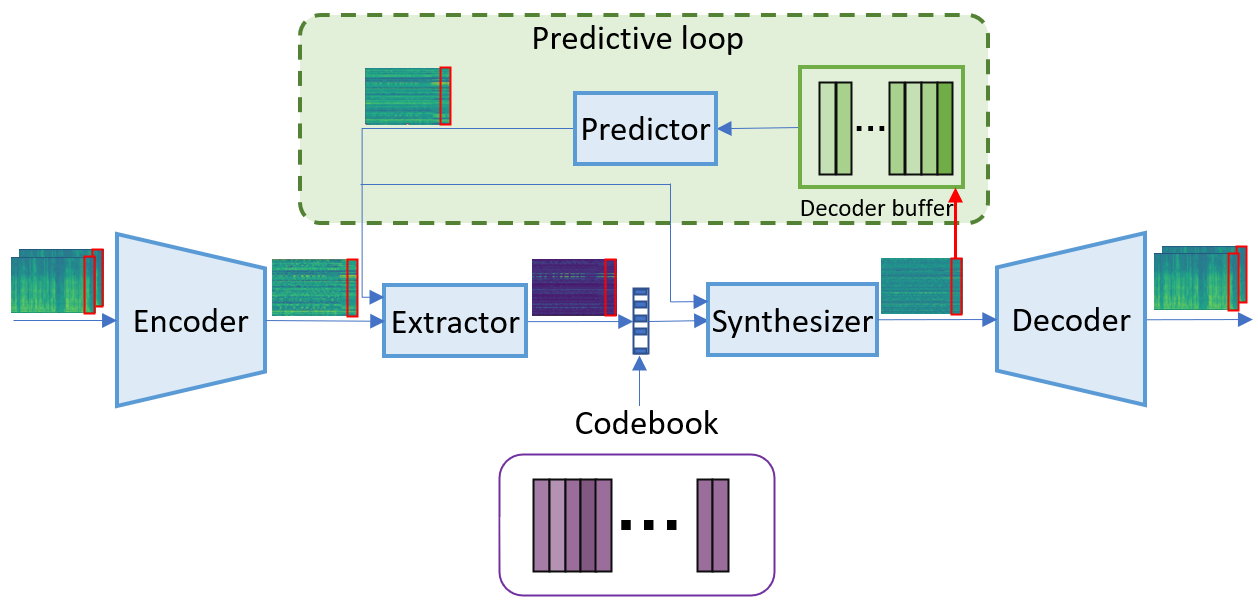}}
  \centerline{(b) Our predictive VQ-VAE based neural codec}\medskip
\end{minipage}
%
\caption{Proposed latent-domain predictive neural speech coding.}
\vspace{-0.4cm}
\label{fig:predictive}
\end{figure}

\IEEEpubidadjcol
Predictive coding is widely used in traditional image \cite{JPEG}, video \cite{H.264/AVC, H.265/HEVC, H.266/VVC} and audio coding \cite{DPCM, ADPCM} for spatial and temporal redundancy removal, where reconstructed neighboring blocks/frames/samples are used to predict the current block/frame/sample and the predicted residuals are quantized and encoded into a bitstream. The residuals after prediction are much sparser and their entropy is largely reduced. In neural video codecs \cite{DVC,DeepContext}, such temporal correlation is also exploited by utilizing motion-aligned reference frame as a prediction or context for encoding current frame. However, in neural audio codec, such techniques have rarely been investigated, to the best of our knowledge. 

Although temporal correlations are exploited in encoder and decoder of neural audio/speech coding by convolutional or recurrent neural networks, these operations can be seen as a kind of open-loop prediction or nonlinear transformation (See Fig. \ref{fig:predictive} (a)). After quantization, the temporal correlation at the decoder side is broken to some extent. We found that for better recovery from quantization noises at low bitrates, the neural network tends to preserve some redundancy in the learned latent representation. Nevertheless, by employing closed-loop prediction as in our predictive coding (see Fig. \ref{fig:predictive} (b)), such a redundancy is eliminated in encoded features but the recovery capability is not affected for closed-loop prediction. The learned latent features are sparse, and the decoding process can achieve high-quality recovery by utilizing the same prediction employed during encoding.

This paper is the first to introduce predictive coding into the VQ-VAE framework for neural speech coding. To reduce the delay, this predictive coding is performed in latent domain as shown in Fig. \ref{fig:predictive} (b). 
Unlike traditional predictive video/audio coding, which subtracts samples from predictions, we introduce a learnable extractor to fuse the prediction with encoder features, obtaining sparse ``new'' information for coding of each frame. All modules are end-to-end learned with adversarial training. Moreover, unlike most previous neural codecs that take time-domain input, we introduce the time-frequency input with a learnable compression on the amplitude. This allows the network to automatically balance the attention paid {to} main and detailed components at different bitrates (see Fig. \ref{fig:predictive} (b)), largely boosting the quality at a bitrate as low as 1 kbps for low-latency speech coding.

The main contributions of this paper are summarized as follows:
\begin{itemize}
\item{We propose TF-Codec, a low-latency neural speech codec that, to the best of our knowledge, is the first real-time codec to report high quality at 1 kbps.}
\item{We introduce predictive coding into the VQ-VAE-based neural speech codec, which largely reduces the temporal redundancy and therefore boosts the coding efficiency.}
\item{We introduce a learnable compression on time-frequency input to adaptively adjust the attention paid {to} main and detailed components at different bitrates.}
\item{We introduce a differentiable vector quantization mechanism based on distance-to-soft mapping and Gumbel-Softmax to facilitate rate control and achieve better rate-distortion optimization.}
\item{We discuss ways to enhance the robustness of predictive neural speech coding under packet losses, which have achieved promising results.}
\end{itemize}

\section{Related Work}
\subsection{Neural Audio/Speech Coding}
\textbf{Generative model-based audio coding} With the advancement of generative models in providing high-quality speech synthesis, researchers have recently proposed leveraging them for speech coding as well \cite{wavcodec,Lyra,sampleRNNcodec,generative,improveopus}, such as WaveNet\cite{wavenet} and LPCNet\cite{LPCNet}. WaveNet was the first to be used as a learned generative decoder to produce high quality audio from a conventional encoder at 2.4 kbps \cite{wavcodec}. Some researchers\cite{improveopus} improved Opus speech quality at low bitrates by using LPCNet for speech synthesis. 
Lyra \cite{Lyra} is a generative model that synthesizes speech from quantized mel-spectrum using an auto-regressive WaveGRU model, producing high-quality speech at 3 kbps. While these methods have achieved good quality at low bitrates, the full potential of neural audio coding has not yet been exploited.

\textbf{End-to-end audio coding} This category learns the encoding, vector quantization, and decoding in an end-to-end manner based on the VQ-VAE\cite{VQ-VAE} framework \cite{VQ-VAE-wavenet,disentangle,soundstream,cascaded,scalable-codec,Configurablecodec,TFNetCodec}. In \cite{VQ-VAE-wavenet}, a VQ-VAE encoder and a WaveNet-based decoder are jointly learned end to end, yielding a high reconstruction quality while passing speech through a compact latent representation corresponding to very low bitrates. The recently proposed SoundStream\cite{soundstream} achieves superior audio quality at a wide range of bitrates from 3 kbps up to 18 kbps with end-to-end learning and a mix of adversarial and reconstruction losses. More recently, an end-to-end audio codec with a cross-module residual coding pipeline was proposed for scalable coding \cite{scalable-codec}. Unlike previous methods based on the waveform input with 1D convolutions, the recent TFNet \cite{TFNetCodec} takes a time-frequency input with a causal 2D encoder-temporal filtering-decoder paradigm for end-to-end speech coding. Among all these methods, the latent features from the encoder are mostly blindly learned without any prior and there are typically temporal correlations remaining in them. In this paper, we propose predictive coding to further remove the redundancies. 

\subsection{Predictive Coding}
\textbf{Classical audio compression}
Predictive coding is widely used in classic audio coding \cite{DPCM, ADPCM, LPC}. As successive audio samples are highly correlated, instead of independently quantizing and transmitting audio samples, the residual between the current sample and its prediction based on past samples is encoded. The DPCM\cite{DPCM}/ADPCM\cite{ADPCM} typically use backward prediction (also known as closed-loop prediction) where past reconstructed samples are used to get the prediction, possibly with some adaptation to the predictor and quantizer in ADPCM.
Another widely used technique in speech coding and processing, linear predictive coding (LPC) \cite{LPC}, leverages a linear predictor to estimate future samples based on the source-filter model. The linear filter coefficients in LPC are computed in an open-loop manner with forward prediction, where original samples rather than reconstructed ones, are used for prediction.
The residuals, along with the linear coefficients, are quantized and encoded.


\textbf{Classical video/image compression} Traditional video coding standards \cite{H.264/AVC, H.265/HEVC, H.266/VVC} always take a predictive coding paradigm for removing temporal redundancies, where a prediction is generated by block-based motion estimation and compensation, and the residual between the original frame and the prediction is transformed, quantized, and entropy coded. In image coding \cite{JPEG} and intra-frame coding of video \cite{H.264/AVC, H.265/HEVC, H.266/VVC}, reconstructed neighboring blocks are used to predict the current block, either in the frequency or pixel domain, and the predicted residuals are encoded. 

\textbf{Deep video compression}
In neural video coding, a typical approach is to replace handcrafted modules, such as motion estimation, with neural networks, while still adhering to a predictive coding paradigm. DVC\cite{DVC} provides more accurate temporal predictions by jointly training motion estimation and compensation networks. The residual information after prediction is then encoded by a residual encoder network. The most relevant work, DCVC\cite{DeepContext}, instead proposes a paradigm shift from predictive coding to conditional coding. It introduces rich temporal context information as a condition for both the encoder and the decoder, largely improving coding efficiency.

{\textbf{Predictive neural speech coding} There have been some attempts in this line of work that are most relevant to our work. The concurrent work \cite{predictivelpc} introduces predictive coding in the parametric domain, where a gated recurrent unit (GRU) based predictor is adopted to predict the LPC coefficients from the past. However, as it is based on LPC analysis under the source-filter assumption, its potential is not fully exploited, and it cannot be easily extended to other signal domains such as music. Another relevant work, the two-stage cognitive coding of speech \cite{cognitivecoding}, leverages predictive coding in the latent space as a representation learning strategy. Different from our motivation, it focuses on representation learning without taking prediction as a coding module.  
}

In light of these methods, we introduce predictive coding into the VQ-VAE framework for neural audio coding, to better remove temporal redundancies and achieve better coding efficiency. {Unlike \cite{predictivelpc}, the predictive coding operates in the latent domain and is trained with other VQ-VAE modules end-to-end, thus extensively exploring its potential.}

\subsection{Autoregressive Model}
Autoregressive generative models have demonstrated strong capabilities in speech synthesis \cite{wavenet,sampleRNN}. They typically generate audio samples one at a time in an autoregressive way, where previously generated samples are used in generating the current sample. Our predictive coding also employs an autoregressive approach, but in contrast to sample-domain autoregression, it operates in the latent domain to reduce the delay by the autoregressive loop. This loop crosses only the quantization layer and does not require passing through the decoder to obtain the output for autoregression.

\subsection{Vector Quantization}
Vector quantization (VQ) is a fundamental technique that is widely used in traditional audio codecs such as Opus \cite{opus} and CELP \cite{celp}. Recently, it has also been applied to discrete representation learning \cite{VQ-VAE} and serves as the basis of end-to-end neural audio coding \cite{VQ-VAE-wavenet,disentangle,soundstream,cascaded,scalable-codec,Configurablecodec,TFNetCodec}. As quantization is inherently non-differentiable, several methods have been proposed in the literature to enable end-to-end learning in neural audio coding, including the one with commitment loss in VQ-VAE\cite{VQ-VAE}, EMA\cite{VQ-VAE}, Gumbel-Softmax based method \cite{Gumbelsoftmax}\cite{vq-wav2vec} and the soft-to-hard technique \cite{soft-to-hard}. VQ-VAE\cite{VQ-VAE} approximates the derivative by the identity function that directly copies gradients from the decoder input to the encoder output. The codebook is learned by moving the selected codeword, found through some distance metric, towards the encoder features. In contrast, the Gumbel-Softmax and soft-to-hard methods introduce the probability of selecting a codeword into the VQ, enabling the selection of discrete codewords in a differentiable way. However, the former employs a linear projection with Gumbel-Softmax to get the probability, which lacks an explicit correlation with quantization error. The latter maps distance to probability and uses soft assignment with annealing during training, potentially leading to a gap between training and inference. Motivated by these works, we propose a Distance-Gumbel-Softmax scheme with rate control that explicitly maps quantization error to probability while uses hard assignment during training and inference.  



\section{The Proposed Scheme}\label{scheme}


\begin{figure*}[ht]
\begin{minipage}[b]{0.5\linewidth}
  \centering
  \centerline{\includegraphics[width=9.2cm]{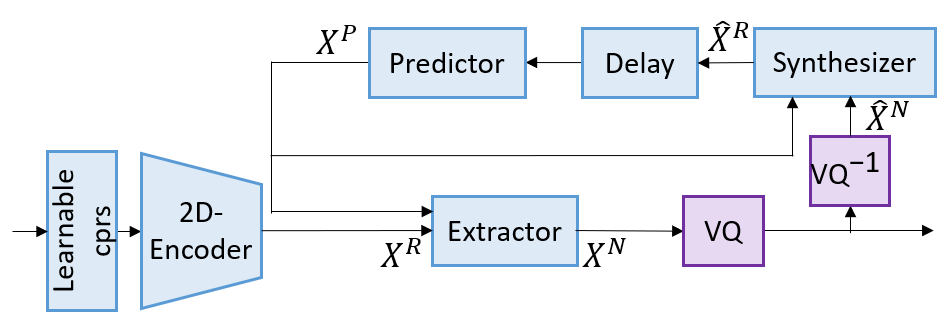}}
 \vspace{-0.2cm}
  \centerline{(a) Encoding}\medskip
\end{minipage}
\hspace{0.5cm}
\begin{minipage}[b]{0.48\linewidth}
  \centering
  \centerline{\includegraphics[width=8cm]{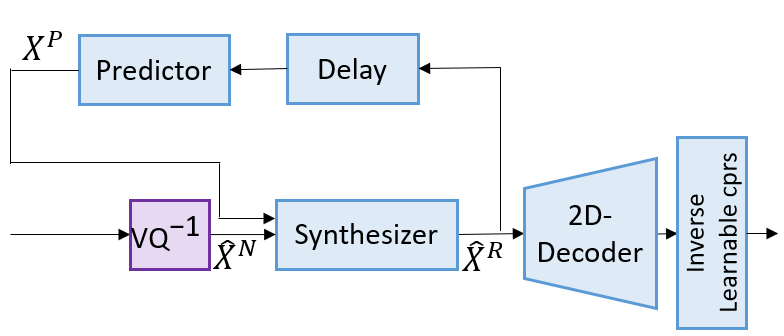}}
   \vspace{-0.2cm}
  \centerline{(b) Decoding}\medskip
\end{minipage}
 \vspace{-0.6cm}
\caption{Encoding and decoding modules for proposed method.}
\vspace{-0.2cm}
\label{fig:encoding-decoding}
\end{figure*}

\subsection{Overview}
Let $x$ denote the time-domain signal to be encoded and $\hat{x}$ the recovered signal after decoding. The optimization of neural audio coding aims to minimize the recovered signal distortion $Dist(x,\hat{x}|\Theta)$ under a given rate constraint, i.e. $R(x|\Theta) \leq R_{target}$. $\Theta$ denotes the neural network parameters. In this paper, we focus on low-latency speech coding.

As shown in Fig. \ref{fig:encoding-decoding}, we employ an encoder to extract latent representations $\bm{X}^{R} = \{\bm{x}_{0}^R,\bm{x}_{1}^R,...,\bm{x}_{T}^R\}$ from $x$, where $\bm{x}_{t}^R$ is the speech feature at frame $t$, and $T$ is the total number of frames.
 For each frame $\bm{x}_{t}^R$ in $\bm{X}^{R}$, a prediction $\bm{x}_{t}^P$ is learned from past reconstructed latent codes  $\hat{\bm{X}}^{R}$ through a predictor $f_{pred}$ with a receptive field of $N$ past frames, given by 
$\bm{x}_t^P = f_{pred}(\hat{\bm{x}}^R_{t-i}\big|{i=1,2,...,N})$. This prediction 
serves as a temporal context for both encoding and decoding. For encoding, an extractor $f_{extr}$ learns residual-like information $\bm{x}_t^N$ from both $\bm{x}_t^R$ and $\bm{x}_t^P$ by $\bm{x}_t^N = f_{extr}(\bm{x}_t^R,\bm{x}_t^P)$, which is ``new'' to past frames. With this autoregressive operation, the temporal redundancy can be effectively reduced. The extracted residual-like feature $\bm{X}^{N}= \{\bm{x}_{0}^N,\bm{x}_{1}^N,...,\bm{x}_{T}^N\}$ is then quantized through a codebook learned by Distance-Gumbel-Softmax and entropy-coded using Huffman coding. For decoding, the quantized residual-like feature $\hat{\bm{x}}_t^N$ in $\hat{\bm{X}}^N$ is merged with the predicition $\bm{x}_t^P$ through a synthesizer $f_{synr}$ to get the current reconstructed latent code $\hat{\bm{x}}_t^R$, given by $\hat{\bm{x}}_t^R = f_{synr}(\hat{\bm{x}}_t^N,\bm{x}_t^P)$. Then, a decoder is employed
to reconstruct the waveform $\hat{x}$. We apply adversarial training to achieve good perceptual quality. In the following subsections, we will describe these techniques in detail.

\subsection{Learnable Input Compression}
The input waveform $x$ is first transformed into the frequency domain with the short-time fourier transform (STFT), yielding a time-frequency spectrum $\bm{X} \in \mathbb{R}^{ 2 \times F\times T}$, where $T$ is the number of frames, $F$ is the number of frequency bins, and $2$ denotes the imaginary and real parts of the complex spectrum. We take frequency domain input instead of the time domain widely adopted in previous works \cite{VQ-VAE,soundstream}, because the frequency domain aligns well with human perception. In this domain, some characteristics of speech (such as the fundamental frequency, harmonics and formants) are explicitly expressed, making it easier for the encoder to learn features related to human perception for coding. As the frequency domain input typically exhibits a high dynamic range and a highly unbalanced distribution due to harmonics, we employ an element-wise power-law compression on the amplitude part given by $|\bm{X}|^p$, where $| \bm{X}|$ is the magnitude spectrum of $\bm{X}$, while the phase is kept unchanged, resulting in the compressed time-frequency spectrum $\bm{X}^{cprs} \in \mathbb{R}^{2\times F\times T}$. The compression acts as a form of input normalization, balancing the importance of different frequencies and ensuring more stable training. Furthermore, we make the power parameter $p$ learnable during training, enabling the model to adapt to different bitrates. To be specific, at low bitrates a higher $p$ may be preferred because it leads to more attention on main components, while at high bitrates, more attention might be paid {to} details with a lower $p$. This technique has proven to be particularly effective for very low bitrate coding, such as 1 kbps, in our experiments.

\subsection{Encoder and Decoder}
\begin{figure}[tb]
\begin{minipage}[c]{1.0\linewidth}
  \centering
  \centerline{\includegraphics[width=7.5cm]{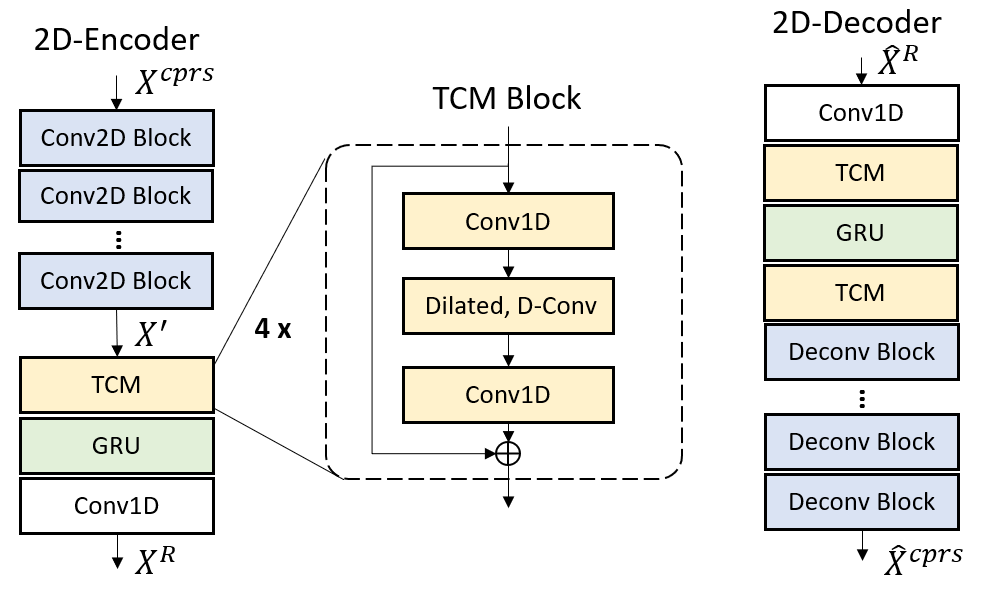}}
\end{minipage}
\caption{Architecture of the encoder and the decoder. D-Conv denotes depthwise convolution.}
\label{fig:enc_dec}
\end{figure}

The encoder takes the compressed time-frequency spectrum $\bm{X}^{cprs} \in \mathbb{R}^{2\times F\times T}$ as input. As shown in Fig. \ref{fig:enc_dec}, five 2D causal convolutional layers are first employed to decorrelate it in two dimensions $(F,T)$ with a kernel size of $(5, 2)$, output channels of $16, 32, 32, 64$ and $64$, and a stride of $1, 2, 4, 4$ and $2$ along the frequency $F$ dimension. The temporal dimension $T$ is kept without any resampling. The output dimension after the five convolutional layers is $C'\times F'\times T $, and we then fold all frequency information into the channel dimension, yielding $\bm{X}' \in\mathbb{R}^{C\times T}, C=C'\times F'$.
 
To capture long-range temporal dependencies, we further employ a temporal convolutional module (TCM) \cite{TCNN} with causal dilated depthwise convolutions followed by a {gated} recurrent unit (GRU) block on $\bm{X}'$, as in \cite{TFNetCodec}, to capture both short-term and long-term temporal dependencies. A final 1D convolutional layer with a kernel size of 1 is used to change the channel dimension to $C_{d}$ for quantization with predictive coding. The encoder finally yields an output $\bm{X}^R=\{\bm{x}_{0}^R,\bm{x}_{1}^R,...,\bm{x}_{T}^R\}\in \mathbb{R}^{C_{d}\times T}$.

The decoder is the opposite of the encoder, reconstructing $\hat{x}$ from the reconstructed features $\hat{\bm{X}}^R=\{\hat{\bm{x}}_{0}^R,\hat{\bm{x}}_{1}^R,...,\hat{\bm{x}}_{T}^R\}\in \mathbb{R}^{C_{d}\times T}$. For better restoration, more TCM modules are used in the decoder than in the encoder. Specifically, one TCM module, one GRU block, and another TCM module are used in an interleaved way to capture local and global temporal dependencies at different depths. Causal deconvolutions are employed to recover the frequency resolution to $F$, and the decoder outputs a feature $\hat{X}^{cprs} \in \mathbb{R}^{2\times F\times T}$. After inverse amplitude compression and an inverse STFT, the waveform $\hat{x}$ is finally reconstructed. The entire process is causal so that it can achieve low latency.

\subsection{Latent-Domain Predictive Coding}\label{predVQ}
As predictive coding is autoregressive, to reduce the delay we only investigate it in the latent domain, as shown in Fig. \ref{fig:predictive} (b) and the encoding/decoding split in Fig. \ref{fig:encoding-decoding}.
\begin{figure}[tb]
\begin{minipage}[c]{1.0\linewidth}
  \centering
   \subfigure[Convolution-based predictor]{\includegraphics[width=1\linewidth]{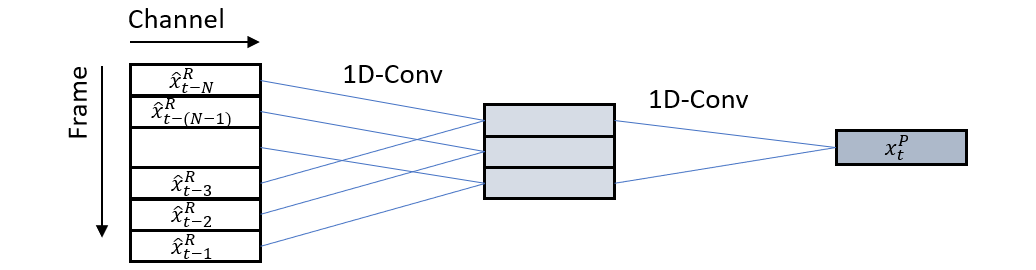}}
    \subfigure[Adaptive predictor]{\includegraphics[width=1\linewidth]{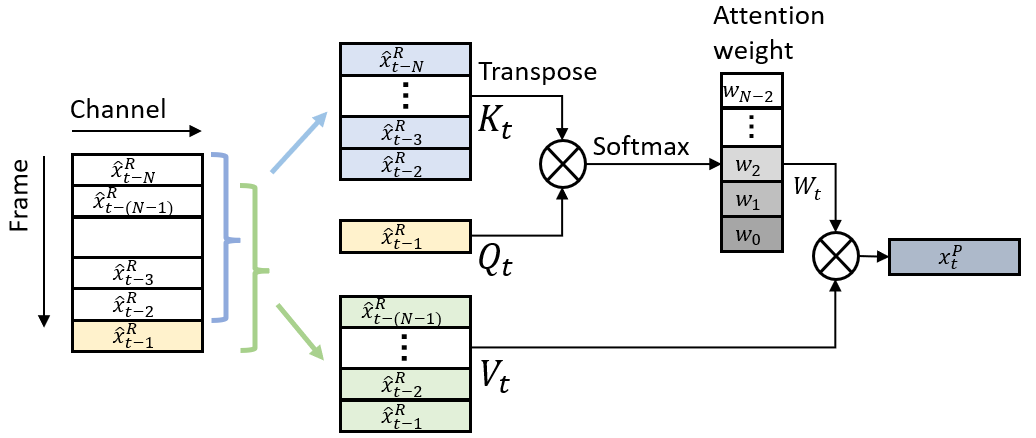}}
\end{minipage}
\caption{The network structure of the predictor.}
\label{fig:predictor}
\end{figure}

\textbf{Predictor} The predictor provides a prediction of the current frame from the past, given by $\bm{x}_t^P = f_{pred}(\hat{\bm{x}}^R_{t-i}\big|{i=1,2,...,N})\in \mathbb{R}^{C_{d}}$, with a window of $N$ frames. As shown in Fig. \ref{fig:predictor}, We investigate two methods for this prediction: (1) Convolution-based predictor, which uses two 1D convolutional layers activated by parametric ReLU (PReLU) \cite{prelu} to achieve a receptive field of 280 ms. 
(2) Adaptive predictor, which learns the prediction kernel from the past to adapt to the time-varying speech signal. The kernel is deduced from the past based on the assumption that the linear prediction coefficients are locally constant.
Specifically, it employs a mechanism similar to self-attention \cite{attention} with the query $\bm{Q}_{t}$, key $\bm{K}_{t}$ and value matrices $\bm{V}_{t}$, defined as follows
\begin{equation}
\left\{\begin{aligned}
\bm{Q}_{t}&=[\hat{\bm{x}}^{R}_{t-1}]^{\mathsf{T}}\in {\mathbb{R}^{1\times C_{d}}}   \\
\bm{K}_{t}&=[\hat{\bm{x}}^{R}_{t-2},\hat{\bm{x}}^{R}_{t-3},...,\hat{\bm{x}}^{R}_{t-N}]^{\mathsf{T}} \in {\mathbb{R}^{(N-1)\times C_{d}}}  \\
\bm{V}_{t}&=[\hat{\bm{x}}^{R}_{t-1},\hat{\bm{x}}^{R}_{t-2},...,\hat{\bm{x}}^{R}_{t-(N-1)}]^{\mathsf{T}} \in {\mathbb{R}^{(N-1)\times C_{d}}}
\end{aligned}
\right.
\end{equation}
It learns an attentive weight matrix $\bm{W}_{t} \in \mathbb{R}^{1 \times (N-1)}$ which {serves} as the prediction kernel by
\begin{equation}
\begin{aligned}
    \bm{W}_t&=Softmax(\bm{Q}_{t}\cdot (\bm{K}_{t})^{\mathsf{T}}/\sqrt{C_{d}}),
\end{aligned}
\end{equation}
where $Softmax(\cdot)$ is the softmax function. The attentive weight matrix is then multiplied with $\bm{V}_{t}$ to get the prediction by [$\bm{x}^P_t]=(\bm{W}^{t} \cdot \bm{V}_t)^{\mathsf{T}}\in \mathbb{R}^{C_{d} \times 1}$. This method is similar to self-attention in how it adaptively captures attention weights from input features, but here we extend it as a kind of prediction. We will show the comparison between these two types of predictors in the experimental section.

To guide the predictor to yield a good temporal prediction for redundancy removal, a prediction loss is introduced in the training as
\begin{equation}
\mathcal{L}_{pred} = \mathbb{E}_x[\frac{1}{C_{d}T}\sum_{t=1}^T||\bm{x}_t^P, sg(\bm{x}_t^R)||_2^2],
\label{eq:pred}\end{equation}
where $sg(\cdot)$ is the stop-gradient operator. At the early training stage, when both the encoder and predictor tend to perform poorly, $\mathcal{L}_{pred}$ forcing $\bm{x}_t^R$ close to $\bm{x}_t^P$ may make the encoder to be confused about generating better representations for reconstruction. Thus, we add the stop-gradient operator on the encoder output $\bm{x}_t^R$ for more stable training and better reconstruction.

\textbf{Extractor and synthesizer} 
Both the extractor $f_{extr}$ and the synthesizer $f_{synr}$ consist of a 1D convolutional layer with a kernel size of 1 and a stride of 1, followed by batch normalizaton (BN) and parametric ReLU as the nonlinear activation function.

\textbf{Training algorithm} As discussed earlier, the predictive loop operates in an autoregressive manner in the latent domain, as the predictor needs the reconstructed context in $\hat{\bm{X}}^R$ for prediction. It is not straightforward to use teacher-forcing for latent-domain autoregression as the vector quantization noise is not easy to model in end-to-end optimization. Instead, we adopt a strategy that facilitates parallel training to improve training efficiency. Specifically, at each iteration, we leverage the model from the previous iteration to get $\hat{\bm{X}}^R$ for prediction in the current iteration. Then the predictive loop can be trained in parallel in the current iteration, thereby speeding up the training process. 

\subsection{Vector Quantization with Rate Control}\label{VQ}
\begin{figure}
\centering
\begin{minipage}[b]{0.48\linewidth}
    {\subfigure[Gumbel-Softmax ]{\includegraphics[width=1\linewidth]{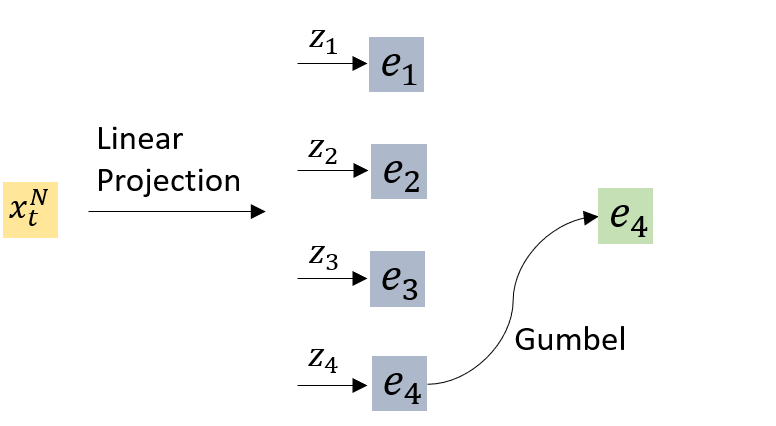}}}
\end{minipage}
\begin{minipage}[b]{0.48\linewidth}
    \subfigure[Distance-Gumbel-Softmax ]{\includegraphics[width=1\linewidth]{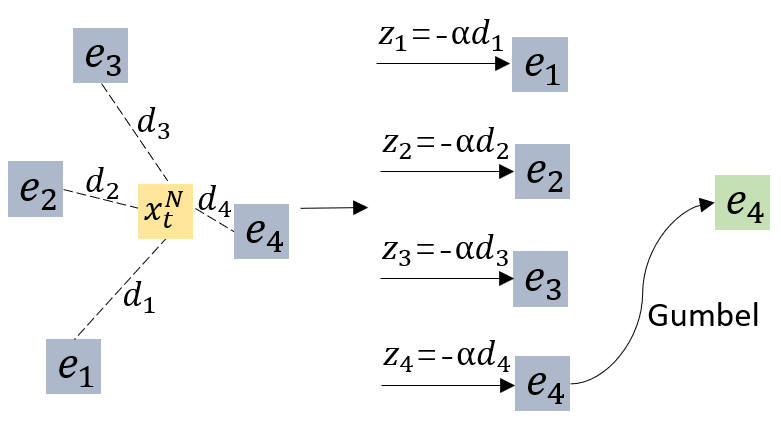}}

\end{minipage}
\caption{Vector quantization mechanism. (a) Gumbel-Softmax in \cite{Gumbelsoftmax}. Latent $\bm{x}_t^N$ is projected to logits $z_i$ through a linear projection and turned into probabilities with Gumbel-Softmax. (b) Our Distance-Gumbel-Softmax. Distance between latent $\bm{x}_t^N$ and codewords $\bm{e}_i$ is first calculated and then mapped to probabilities with Gumbel-Softmax. }
\label{fig:gumbel}
\end{figure}

As discussed in Section II.D, Gumbel-Softmax \cite{Gumbelsoftmax}\cite{vq-wav2vec} and soft-to-hard \cite{soft-to-hard} methods introduce the probability of selecting a codeword, thereby making rate control feasible. However, Gumbel-Softmax employs a linear projection to select the codeword without explicitly correlating it with the quantization error, as illustrated in Fig. \ref{fig:gumbel} 
 (a). 
The weighted average operation in the soft-to-hard method can easily lead to a gap between training and inference. 
In light of these limitations, we propose a Distance-Gumbel-Softmax method, as shown in Fig. \ref{fig:gumbel} (b), for quantization, which leverages the advantages of both methods to provide a quantization-error-aware assignment with rate control. 

\textbf{Distance-Gumbel-Softmax-based VQ}
As shown in Fig. \ref{fig:gumbel} (b), given a codebook with $K$ codewords $E = \{\bm{e}_1,\bm{e}_2,...,\bm{e}_K\} \in {\mathbb{R}^{C_{d}\times K}}$, we first compute the distance between the current latent vector $\bm{x}_t^N \in\mathbb{R}^{C_{d}}$ and all $K$ codewords as
\begin{equation}
\begin{aligned}
    \bm{d}_{t} &=[\ell_d(\bm{x}_t^N, \bm{e}_1),...,\ell_d(\bm{x}_t^N, \bm{e}_K)] \in \mathbb{R}^{K},
\end{aligned}
\end{equation}
where $\ell_d$ is a distance metric, and we use $\ell_2$ in our implementation. Then, the distance is mapped to
logits $\bm{z}_t$, given by $\bm{z}_t = -\alpha\cdot\bm{d}_t$, where $\alpha$ is a positive scalar to control the mapping from distance $\ell_d(\bm{x}_t^N, \bm{e}_k)$ to logits $z_{t,k}$ such that the codeword closer to the current feature $\bm{x}_t^N$ will have a higher probability of being selected, and we set $\alpha$ to $5$. Then, we employ the Gumbel-Softmax to get the probability $\bm{\mu}_t$ for codebook assignment, given by $\bm{\mu}_{t} =GumbelSoftmax(\bm{z}_t) \in \mathbb{R}^{K}.$
Thus, the probability for selecting the $k$-th codeword $\bm{e}_k$ to quantize $\bm{x}_t^n$ is given by
\begin{equation}
\begin{aligned}
    \mu_{t,k}
    &=\frac{exp((-\alpha \cdot d_{t,k} + v_{t,k})/\tau)}{\sum_{i=1}^K exp((-\alpha \cdot d_{t,i} + v_{t,i})/\tau)},
\end{aligned}
\end{equation}
where $\tau$ is the temperature of Gumbel-Softmax, which is exponentially annealed from $2$ to $0.5$ in our experiment, and $v_{t,k} \sim Gumbel(0,1)$ are samples drawn from the Gumbel distribution. During the forward pass, the hard index $argmax_{k\in\{1,2,..,K\}} \mu_{t,k}$ is selected; thus, there is no gap between training and inference. During the backward pass, the gradient with respect to logits $\bm{z}_t$ is used.

\textbf{Entropy estimation and rate control}
As entropy serves as the lower bound of actual bitrates, we leverage entropy estimation to control the bitrate $R(x|\Theta)$ towards a given target $R_{target}$, motivated by the work in \cite{soft-to-hard, scalable-codec}. Using the Distance-Gumbel-Softmax based VQ, we can calculate the sample soft assignment distribution $\tilde{\bm{\mu}}$, by summing up the Softmax probabilities to each codeword within a minibatch $\tilde{\mu}_{t,k,b}$ as
\begin{equation}
\begin{aligned}
    {\tilde{\mu}}_k &= \frac{1}{BT}\sum_{b=1}^B \sum_{t=1}^T \tilde{\mu}_{t,k,b}, k \in\{1,2,...,K\},
\end{aligned}
\end{equation}
where $B$ and $T$ represent the batch size and the number of frames in each audio clip, respectively, and $\tilde{\mu}_{t,k,b}$ is the Softmax distribution probabilities given by $\tilde{\mu}_{t,k,b} = Softmax(z_{t,k,b})$ over $K$ codebook entries. Then, we can estimate the ``soft entropy" on the soft assignment distribution $\tilde{\bm{\mu}} \in \mathbb{R}^K$ as 
\begin{equation}
\begin{aligned}
    \mathcal{H}(\tilde{\bm{\mu}}) &\approx -\sum_{k=1}^K \tilde{\mu}_k\log \tilde{\mu}_k.
\end{aligned}
\end{equation}

The rate control is conducted over each minibatch with the following loss function $\mathcal{L}_{rate}$:
\begin{equation}
\begin{aligned}
\mathcal{L}_{rate} &= |R_{target} - \mathcal{H}(\tilde{\bm{\mu}})|.
\end{aligned}
\label{eq:L-rate}
\end{equation}
This loss $\mathcal{L}_{rate}$ not only constraints the bitrate but also performs rate-distortion optimization by $\mathcal{L}_{RD} = Dist(x, \hat{x}) + \lambda \cdot \mathcal{L}_{rate}$. When the current entropy is higher than $R_{target}$, it will push similar features to be quantized to the same codeword through a trade-off between rate and distortion; whereas when it is lower than $R_{target}$, similar features may be quantized to different codewords to retain higher quality but at a higher rate. It should be noted that although there are some estimations here, we found that the actual bitrate is controlled well during testing.

To reduce the codebook size for easy optimization, group vector quantization\cite{vq-wav2vec} is employed. Specifically, each frame $\bm{x}_t^N\in \mathbb{R}^{C_d}$ is split into $G$ groups along the channel dimension, yielding  $\bm{x}_t^{'N} \in \mathbb{R}^{ \frac{C_d}{G}\times G}$, and each group is quantized with a separate codebook containing $K$ codewords $\{e'_1,e'_2,...,e'_K\} \in \mathbb{R}^{\frac{C_d}{G}\times K}$.  Moreover, four overlapped frames (40 ms new data) are merged into one for quantization so the codeword dimension is $\frac{C_d}{G}\times4$ for each codebook. Unlike common codebook settings in existing neural codecs\cite{soundstream}\cite{encodec}, we establish a larger codebook size so that it can capture the real distribution of the latent features through the rate-distortion optimization. For example, at 3 kbps, each 40 ms new data is expected to consume 120 bits. The codebook parameters $G$ and $K$ are set to 16 and 1024, respectively, where $G \cdot \log_2(K) = 16 \cdot \log_2(1024) = 160 > 120$. The real bitrate is then controlled by Eq. \ref{eq:L-rate} to achieve 3 kbps. This is quite different from the diversity loss in Gumbel-Softmax based method \cite{Gumbelsoftmax}, where a uniform distribution is enforced on the codeword usage. Table \ref{table:codebook} shows the codebook configurations $G$ and $K$ at various bitrates in our experiment.

\subsection{Adversarial Training}
Adversarial training has been shown to be very effective in high-quality speech generation \cite{hifigan}\cite{melgan}. For high reconstructed perceptual quality, we also employ adversarial training in our scheme with a frequency-domain discriminator. It takes the complex time-frequency spectrum of the input waveform as input. The magnitude spectrum is power-law compressed with a power of 0.3 {to balance the relative importance of different components. The phase is kept unchanged.} 
Four 2D convolutional layers with a kernel size of $(2,3)$ and a stride of $(2,2)$ are used to extract features with progressively reduced resolutions in both frequency and time dimensions. The channel numbers are 8, 8, 16 and 16, respectively. Each convolutional layer is followed by an instance normalization (IN) and a Leaky ReLu \cite{leaklyrelu}. A linear transformation is used to fold all frequency information into channels and reduce the channel dimension to 1. Finally, we use a temporal average pooling layer with a kernel size of 10 and produce the final downsampled one-dimensional logits of size $T_d$ in the time dimension.

We use the least-square loss as the adversarial objective, similar to that in LSGAN\cite{lsgan}. 
The adversarial loss for the generator $G$ is
    \begin{equation}
        \mathcal{L}_{adv}=\mathbb{E}_{\hat{x}}[\frac{1}{T_d}\sum_{t=1}^{T_d}(D_t(\hat{x})-1)^2].
        \label{eq:adv}
    \end{equation}
where $\hat{x}=G(x)$ is the reconstructed signal. The loss for the discriminator $D$ is
    \begin{equation}
    \mathcal{L}_{D}=\mathbb{E}_x[\frac{1}{T_d}\sum_{t=1}^{T_d}(D_t(x)-1)^2]+\mathbb{E}_{\hat{x}}[\frac{1}{T_d}\sum_{t=1}^{T_d}(D_t(\hat{x}))^2].
        \label{eq:discrim}
    \end{equation}
We also employ a feature loss $\mathcal{L}_{feat}$ to guide the training of the generator for high perceptual quality \cite{hifigan}\cite{melgan}. It is computed as the $\ell_1$ difference of the deep
features from the discriminator between the generated and the original audios, given by
    \begin{equation}
        \mathcal{L}_{feat}=\mathbb{E}_x[\frac{1}{L}\sum_{l=1}^L{\frac{1}{C_{l}F_{l}T_{l}}}||D^{l}(x)-D^{l}(\hat{x})||_1],
        \label{eq:feat}
    \end{equation}
where $D^{l} ,l\in\{1,2,...,L\} $ is the feature map of the $l$-th layer of the discriminator, and $C_l,F_l,T_l$ denotes the channel, frequency and time resolutions of $D^{l}$, respectively. We compute the feature loss on the first four 2D convolutional layers of the discriminator. 


\subsection{Objective Function}
We employ the following loss function to guide the training for maximized output audio quality at the target bitrate. The total loss for the generator consists of a reconstruction term $\mathcal{L}_{recon}$, a rate-constraint term $\mathcal{L}_{rate}$, a prediction term $\mathcal{L}_{pred}$, an adversarial term $\mathcal{L}_{adv}$, and a feature-matching term $\mathcal{L}_{feat}$, i.e.
\begin{equation}
\label{eq:loss}
\begin{aligned}
\mathcal{L}_G = \mathcal{L}_{recon} + \lambda_{rate} \mathcal{L}_{rate} + \lambda_{pred}\mathcal{L}_{pred} \\ + \lambda_{adv}\mathcal{L}_{adv} + \lambda_{feat}\mathcal{L}_{feat},
\end{aligned}
\end{equation}
where $\mathcal{L}_{rate}$, $\mathcal{L}_{pred}$, $\mathcal{L}_{adv}$ and $\mathcal{L}_{feat}$ have been explained in Eq. \ref{eq:L-rate}, \ref{eq:pred}, \ref{eq:adv} and \ref{eq:feat}, respectively. The reconstruction term is selected to achieve both high signal fidelity and high perceptual quality. We use two frequency-domain terms for $\mathcal{L}_{recon}$, as shown below
    \begin{equation}
    \mathcal{L}_{recon} =\mathcal{L}_{bin} + \lambda_{mel}\mathcal{L}_{mel}.
    \end{equation}
The first term $\mathcal{L}_{bin}$ is the mean-square-error (MSE) loss on the power-law compressed STFT spectrum \cite{powermse}. To maintain STFT consistency \cite{stft-consistency}, the reconstructed spectrum is first transformed to the time domain and then to the frequency domain to calculate the loss. The second term $\mathcal{L}_{mel}$ is the multi-scale mel-spectrum loss given by
    \begin{equation}
        \mathcal{L}_{mel}=\mathbb{E}_x[\frac{1}{W}\sum_{n=1}^W{\frac{1}{T_{n}S}}||\phi^{n}(x)-\phi^{n}(\hat{x})||_{1}],
    \end{equation}
where $\phi^{n}(\cdot)$ is the function that transforms a waveform into the mel-spectrogram using the $n$-th window size $T_{n}$. $S$ denotes the number of mel bins, which is set to 64 for all window lengths. Following \cite{melloss}, we calculate the mel spectra over a sequence of window-lengths between 64 and 2048. 
We set $\lambda_{rate}=0.04$, $\lambda_{pred}=0.02$, $\lambda_{adv}=0.001$, $\lambda_{feat}=0.1$, and $\lambda_{mel}=0.25$ to balance different terms in our experiments.


\begin{figure*}[tb]
\centering
\begin{minipage}{1\linewidth}
\centering
\hspace{-0.4cm}
\subfigcapskip=-3pt
\subfigure[TF-Codec vs. standard codecs]{\includegraphics[width=4.8cm]{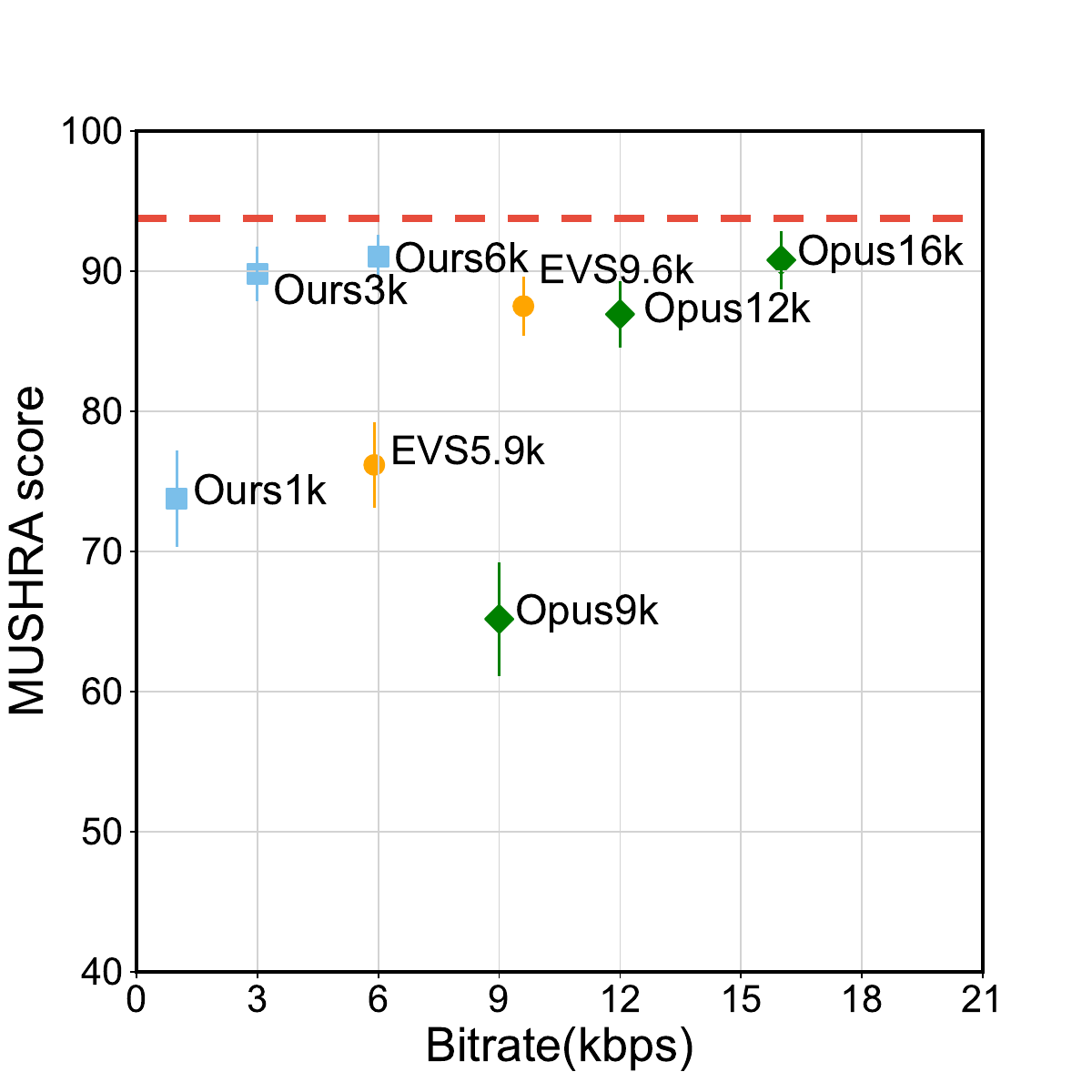}}
\hspace{-0.5cm}
\subfigcapskip=-3pt
\subfigure[TF-Codec vs. neural codecs]{\includegraphics[width=4.8cm]{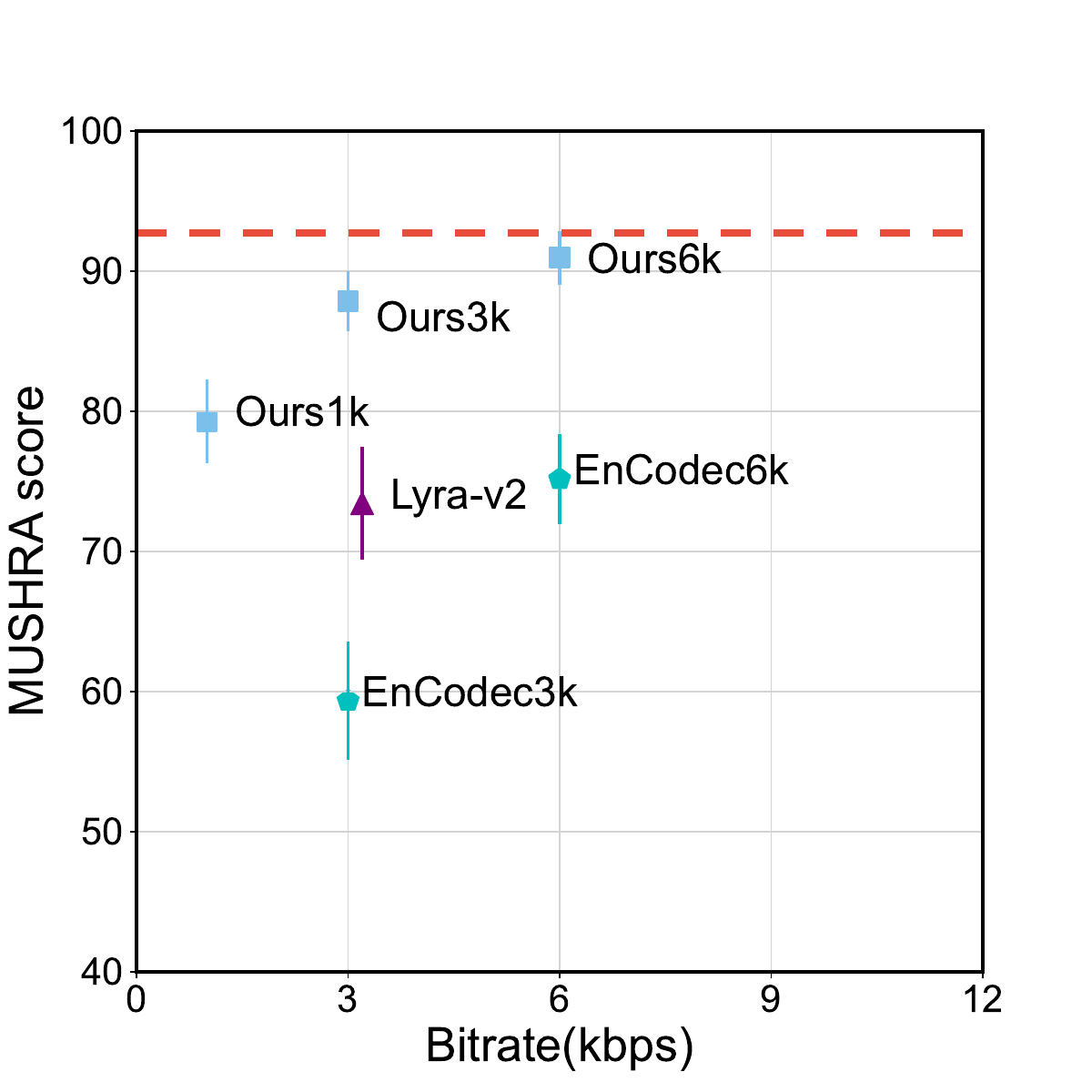}}
\hspace{-0.5cm}
\subfigcapskip=-3pt
\subfigure[Entropy]{\includegraphics[width=4.8cm]{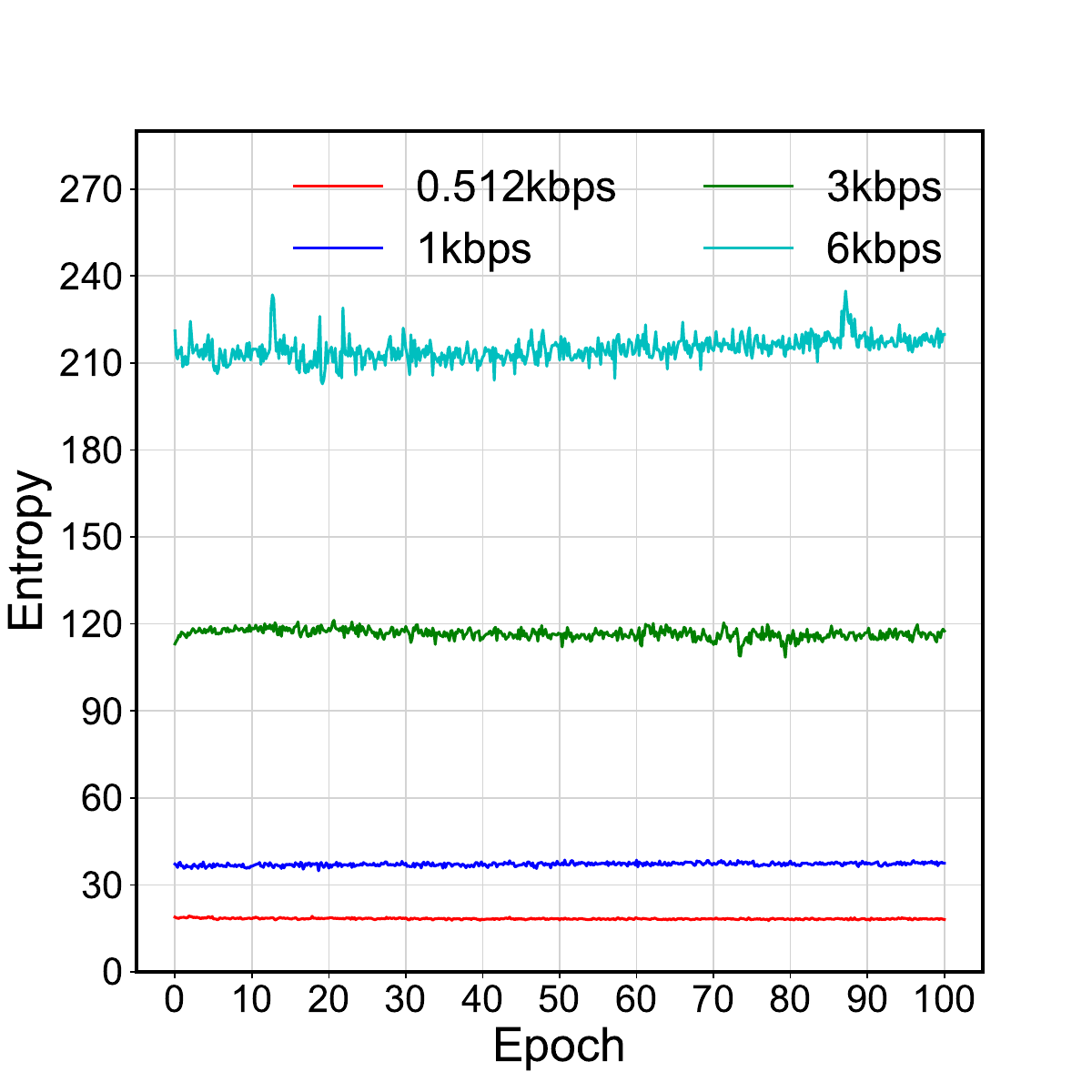}}
\hspace{-0.5cm}
\subfigcapskip=-3pt
\subfigure[The learnable power]{\includegraphics[width=4.8cm]{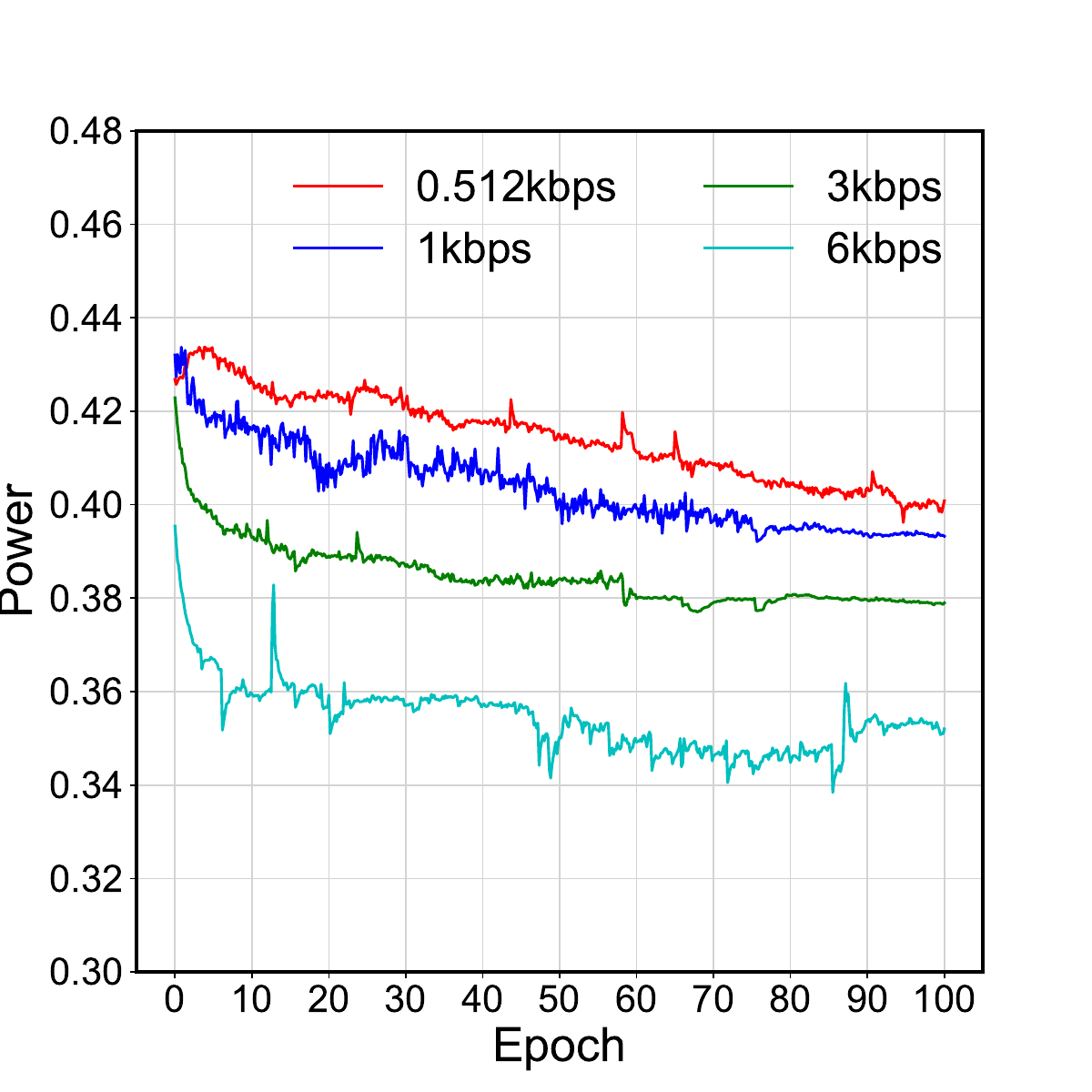}}
      
      
      
\end{minipage}
\vspace{-0.4cm}
\caption{(a)(b) Subjective evaluation results. The red dotted line represents the score of the reference.  The error bar denotes $95\%$ confidence intervals. We use Opus-1.3.1 and EVS-16.2.0 from the official release and set them to WB mode. Audios are sampled at 16 kHz for these codecs, except for Encodec which operates at 24 kHz. For Encodec, the 16khz audio is upsampled to 24 kHz. The convolution-based predictor is used for TF-Codec in this test. {It's worth noting that  TF-Codec and Opus operate in a variable bitrate (VBR) manner. Both Lyra-v2 and Encodec operate in constant bitrate (CBR) mode in this figure. EVS is set to source-controlled variable bitrate (SC-VBR) \cite{evs} mode at 5.9 kbps and CBR at 9.6 kbps.}
(c) Entropy for 40ms data during training. (d) The learned power coefficient during training. In (c) and (d), the curves correspond to the adversarial training stage only.}

\vspace{-0.2cm}
\label{fig:evaluation}
\vspace{-0.3cm}
\end{figure*} 

\section{Experimental Results}\label{EXP}
In this section, we evaluate the proposed TF-Codec against the state-of-the-art and provide a detailed analysis of each part to demonstrate what it learns and why it works effectively.

\subsection{Datasets and Settings}
We take 890 hours of 16kHz clean speech from the Deep Noise Suppression Challenge at ICASSP 2021 \cite{DNSchallenge}, which includes multilingual speech, emotional, and singing clips. Each audio is cut into 3-second clips with a random speech level from $[-50, -10]$ db for training. For evaluation, we use 1458 clips of 10 seconds without any overlap with the training data, covering more than 1000 speakers with multiple languages. A Hanning window is used in STFT with a window length of 40 ms and a hop length of 10 ms.

All the modules of the TF-Codec, including the encoding, decoding, and quantization, can be trained end-to-end in a single stage. For adversarial training, we first train a good generator end-to-end and then fine-tune the generator with a discriminator in an adversarial manner.

During training, we use the Adam optimizer\cite{adam} with a learning rate of $3\times10^{-4}$ for the generator in the first stage. Then, the generator and discriminator are trained with a learning rate of $3\times10^{-5}$ and $3\times10^{-4}$, respectively. We train both stages for 100 epochs with a batch size of 100. 

\begin{table}
\centering
\caption{Codebook configurations and bitrate analysis on test set.\label{table:codebook}}
\begin{tabular}{ccc}
\toprule
\multirow{2}{*}
{\makecell[c]{\\Bitrate Modes\\(kbps)}} & Codebook size  & {Huffman coding} \\ 
\cmidrule(lr){2-2}\cmidrule(lr){3-3}
& $G\cdot\log_2(K)$ & \makecell[c]{Average bitrate \\ (kbps)} \\ 
\midrule
0.512          & G=3, K=512        & \multicolumn{1}{c}{0.498 }          \\ 
1              & G=6, K=1024       & \multicolumn{1}{c}{1.014 }          \\ 
3              & G=16, K=1024      & \multicolumn{1}{c}{3.089 }       \\ 
6              & G=32, K=1024      & \multicolumn{1}{c}{6.162 }     \\ 
\bottomrule
\end{tabular}
\end{table}


\subsection{Comparison with State-of-the-Art Codecs}
We first compare the proposed TF-Codec with several traditional codecs and two latest neural codecs to demonstrate the strong representation capability of our backbone. We conduct a subjective listening test using a MUSHRA-inspired crowd-sourced method \cite{mushra}, where 10 participants evaluate 15 samples from the test set. In the MUSHRA evaluation, the listener is presented with a hidden reference and a set of test samples generated by different methods. The low-pass-filtered anchor is not used in our experiment.

 We take two standard codecs, i.e., Opus and EVS. for comparison.
 Opus\footnote{https://opus-codec.org} \cite{opus} is a versatile codec widely used for real-time communications, supporting narrowband to fullband speech and audio with a bitrate from 6 kbps to 510 kbps. The EVS codec\cite{evs} is developed and standardized by 3GPP primarily for Voice over LTE (VoLTE). We also compare with two latest neural codecs, i.e., Lyra-v2 and the concurrent work Encodec \cite{encodec}. Lyra-v2\footnote{https://github.com/google/lyra} is an improved version of Lyra-v1, which {integrates} SoundStream's architecture\cite{soundstream} for better audio quality and bitrate scalability\footnote{https://opensource.googleblog.com/2022/09/lyra-v2-a-better-faster-and-more-versatile-speech-codec.html}. Encodec\footnote{https://github.com/facebookresearch/encodec} produces high-fidelity audio across a wide range of bandwidths and audio types.
 

 Fig. \ref{fig:evaluation} (a)(b) shows the evaluation results, where we compare our TF-Codec from 1 kbps to 6 kbps against standard and neural codec baselines at various bitrates.
 It is observed that our TF-Codec at 1 kbps significantly outperforms Opus at 9 kbps, Lyra-v2 at 3.2 kbps, and EnCodec at 6 kbps, demonstrating the strong representation capability of the TF-Codec. When operating at 3 kbps, our TF-Codec achieves better performance than EVS at 9.6 kbps and Opus at 12 kbps, and also clearly outperforms two neural codecs. In the higher bitrate range, our TF-Codec at 6 kbps performs on par with Opus at 16 kbps and outperforms Encodec at 6 kbps by a large margin. Besides, Fig. \ref{fig:evaluation} (c) shows that the total entropy of our TF-Codec {is kept under control} with the rate loss $\mathcal{L}_{rate}$ during training. Table \ref{table:codebook} displays the actual bitrate after Huffman coding on the test set. We can see that the actual bitrates are well controlled. As variable-length coding cannot ensure constant bits for each frame, we also collect the statistics of per-frame bit consumption. At 3kbps, the maximum, minimum, and average bits across all frames of the entire test set is 190, 39 and 124, respectively. For a single audio, the variation is smaller with an average standard deviation of 40 and a maximum of 53.



\subsection{Ablation Study}
To evaluate different parts of the proposed method, we employ several objective metrics, including wideband PESQ \cite{pesq}, STOI \cite{stoi}, and VISQOL \cite{visqol}. Although these metrics are not designed and optimized for precisely the same task, we found that for the same kind of distortions in all compared schemes, they align well with perceptual quality. 

\begin{table}[!t]
\begin{center}
\caption{Evaluation on predictive coding at 3 kbps(w/o. adv).\label{table:contextual_coding}}
\vspace{-0.2cm}
\begin{tabular}{lccc}
\toprule
 Methods & PESQ & STOI & ViSQOL \\
\midrule
TF-Codec w/o. Prediction  & 2.763 & \textbf{0.917} & 3.219\\
TF-Codec w. Adapt & 2.774 & 0.914 &  3.332\\
TF-Codec w. Conv & \textbf{2.895} & \textbf{0.917} & \textbf{3.345}\\
\bottomrule
\end{tabular}
\end{center}
\vspace{-0.2cm}
\end{table}
\begin{figure}
\centering
\begin{minipage}[b]{0.55\linewidth}
    {\subfigure[Feature visualization]{\includegraphics[width=1\linewidth,height=1\linewidth]{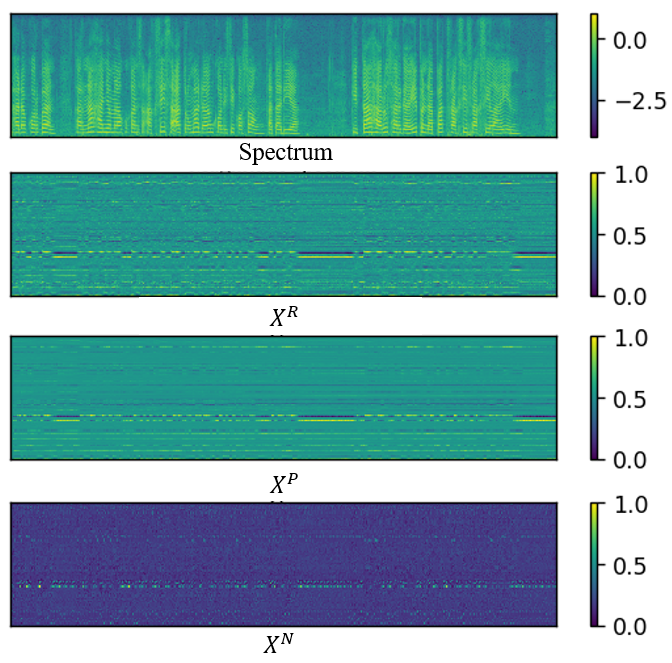}}}
\end{minipage}
\begin{minipage}[b]{0.4\linewidth}
    {\subfigure[Non-predictive]{\includegraphics[width=0.84\linewidth]{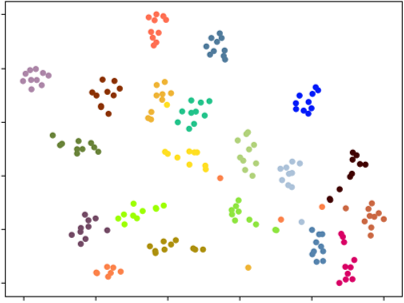}}}
    {\subfigure[Predictive]{\includegraphics[width=0.85\linewidth]{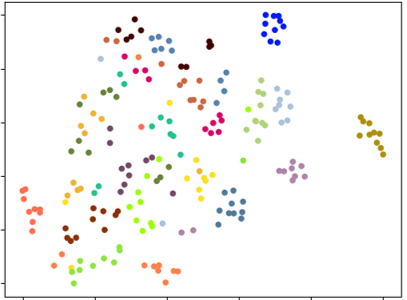}}}
\end{minipage}
\caption{Feature visualization of predictive coding. (a) The four rows from top to bottom show the STFT spectrum (log-scale) of the uncompressed audio, the output of the encoder before predictive coding, the output of the predictor, and that of the extractor, respectively. Values are linearly normalized between 0 and 1. (b)(c) T-SNE visualization of speaker information by non-predictive and predictive coding at 3 kbps, respectively.}
\label{fig:contextual visual}
\end{figure}
\subsubsection{Predictive coding}


We first evaluate the effectiveness of predictive coding. We compare the two variations of the TF-Codec by convolution-based and adaptive predictors, respectively, with the one without predictive coding by disabling the predictive loop. Table \ref{table:contextual_coding} shows the evaluation results where the adversarial training is disabled for all compared methods. It can be seen that when all operating at 3kbps, the predictive coding improves the reconstructed audio quality in both PESQ and ViSQOL with similar speech intelligibility as measured by STOI. The convolution-based method outperforms the adaptive mechanism because, after quantization, the assumption for local constant linear prediction may not hold anymore in the adaptive scheme. 

To further look into the representations it learns, we visualize the features of different modules by predictive coding in Fig. \ref{fig:contextual visual} (a). The four rows from top to bottom show the STFT spectrum of the uncompressed audio, the output of the encoder $\bm{X}^R$ before predictive coding, the output of the predictor $\bm{X}^P$, and that of the extractor $\bm{X}^N$. It can be observed that the prediction $\bm{X}^P$ is quite similar to $\bm{X}^R$, indicating that the predictor provides a good prediction of the current frame from the past. We can also observe that the feature $\bm{X}^N$ after the extractor becomes much sparser than $\bm{X}^R$, indicating that most redundant information has been removed. We also calculate the temporal correlation coefficient of the learned representation, i.e., the last layer output of the encoding. The non-predictive coding without the predictive loop achieves an average correlation coefficient of 0.37, while predictive coding reduces it to 0.09, showing that the temporal correlation is removed more thoroughly in predictive coding. This is also consistent with the visualization results in Fig. \ref{fig:contextual visual}(a).

To further explore what redundant information has been removed by predictive coding, we show the t-SNE \cite{tsne} visualization of the speaker information contained in the learned representations in Fig. \ref{fig:contextual visual}(b)(c) for non-predictive and predictive coding, respectively. To achieve this, we perform temporal pooling on the learned representation, yielding one embedding vector for each audio. The utterances of 20 randomly selected speakers from the Librispeech dataset \cite{librispeech} are used for visualization. We can observe that representations from non-predictive coding are well clustered for each speaker, showing that they contain most speaker information. In contrast, in predictive coding, the embeddings scatter for most speakers indicating that the speaker information is effectively removed. This is reasonable, as speaker-related information is relatively constant in time and easy to predict.



\subsubsection{Learnable input compression}
To demonstrate the effectiveness of the learnable input compression, we compare it with a fixed power-law compression where the power parameter is set to 0.3 as that in \cite{powermse}. Table \ref{table:learnable_input} shows that at 1 kbps, the learnable compresssion clearly outperforms the fixed one in all three metrics, exhibiting both better perceptual quality and better speech intelligibility. In our subjective evaluation, we also found obvious perceptual quality boost for very low bitrates, such as 1 kbps. To investigate what power parameters it learns, we also report the learned power $p$ at various bitrates during the training, as shown in Fig. \ref{fig:evaluation} (d). We can see that the learned power gradually decreases during the training process, indicating that the model first mainly focuses on high-energy bins, usually the low-frequency bands. As the epoch increases, the model turns to pay more attention to the low-energy details of the spectrum. It is also observed that the higher the bitrate, the smaller the $p$, which means that the model tries to examine the detailed components with more bits available, yielding better perceptual quality.

\begin{table}[!t]
\begin{center}
\caption{Evaluation on learnable input compression at 1 kbps.\label{table:learnable_input}}
\vspace{-0.2cm}
\begin{tabular}{lccc}
\toprule
 Methods & PESQ & STOI & ViSQOL\\
\midrule 
fixed compression  & 2.289 & 0.877 & 2.781\\
learnable compression  & \textbf{2.351} & \textbf{0.887} & \textbf{2.851}\\
\bottomrule
\end{tabular}
\end{center}
\vspace{-0.2cm}
\end{table}

\subsubsection{Distance-Gumbel-Softmax-based VQ}
We compare the Distance-Gumbel-Softmax-based VQ mechanism with the previous Gumbel-Softmax-based method in \cite{Gumbelsoftmax}. Table \ref{table:vq} shows that at 3 kbps, our method outperforms the previous Gumbel-Softmax-based method in all metrics, indicating that the explict injection of distance information helps improve reconstruction quality.

\begin{figure}[tb]
\begin{minipage}[c]{1.0\linewidth}
  \centering
  \hspace{-0.4cm}
   \subfigure[Codeword usage histogram]{\includegraphics[width=1\linewidth,height=0.4\linewidth]{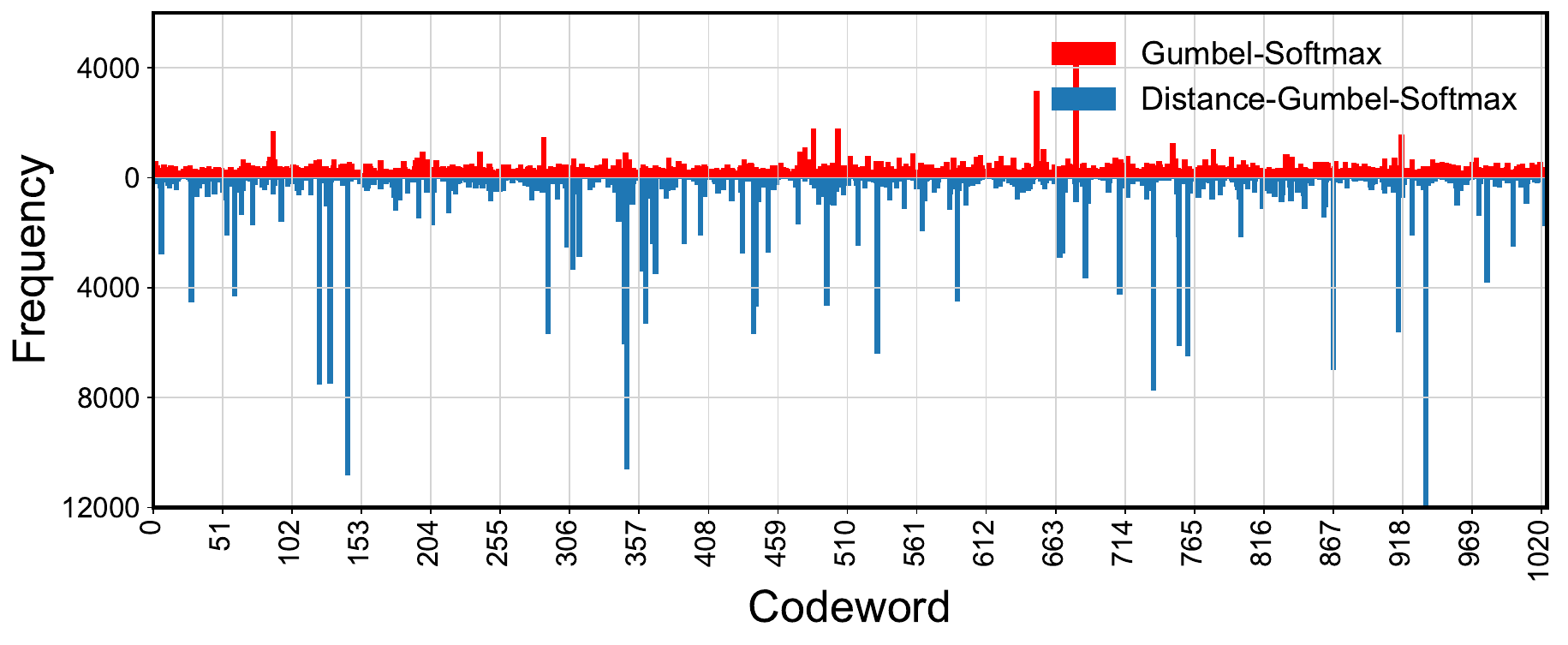}}
    \subfigure[Codeword confidence]{\includegraphics[width=1\linewidth,height=0.4\linewidth]{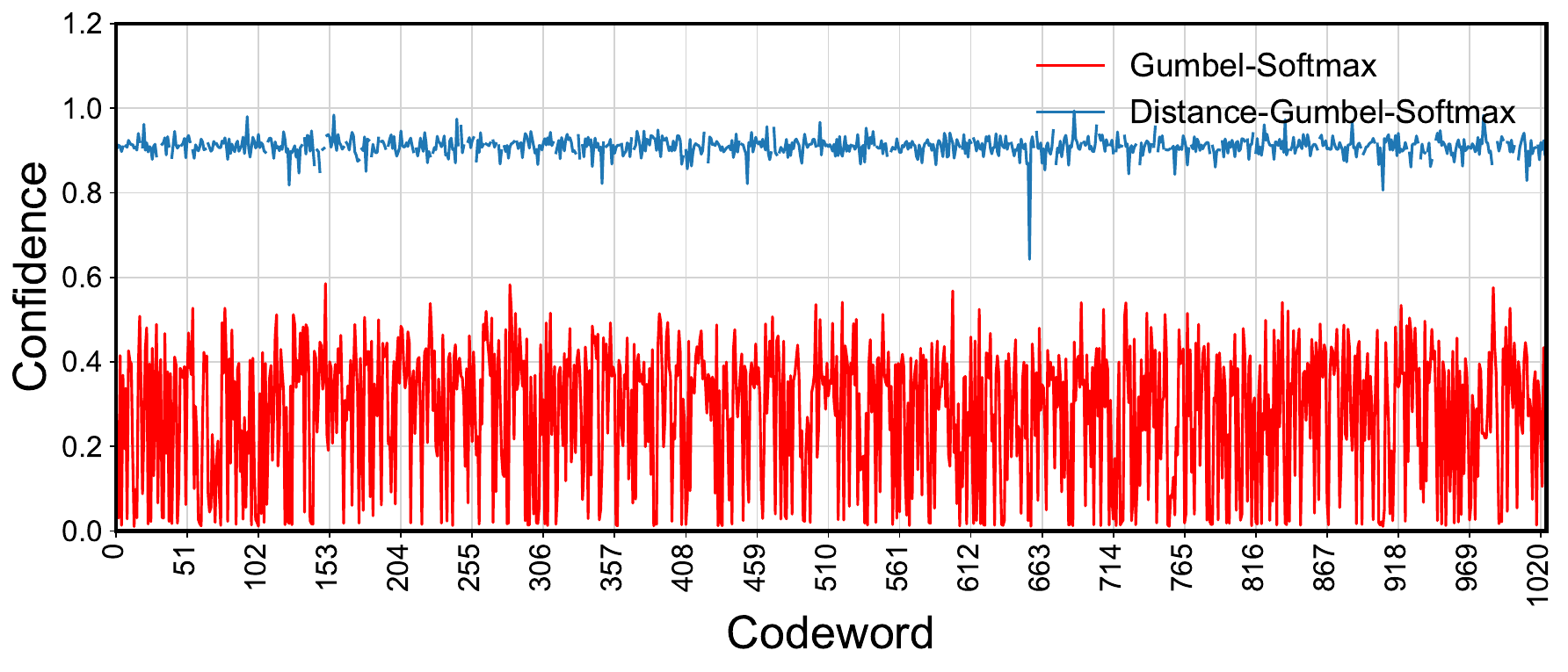}}
\end{minipage}
\caption{Characteristics of one randomly selected learned codebook at 3 kbps. (a) The frequency of 1024 codewords being selected. (b) The confidence score of 1024 codewords.}
\label{fig:codeword}
\end{figure}

We also show the distribution of the learned codebooks to help understand how the Distance-Gumbel-Softmax-based vector quantization learns. Fig. \ref{fig:codeword}(a) shows the usage of 1024 codewords of one codebook on the test set with 1458 audios for both the Gumbel-Softmax-based method \cite{Gumbelsoftmax} and the proposed Distance-Gumbel-Softmax mechanism. We can observe that the codewords tend to be more uniformly distributed in the Gumbel-Softmax-based method, while in the Distance-Gumbel-Softmax, the codewords are distributed more diversely, with some codewords being used very frequently. This is reasonable, as in the Gumbel-Softmax-based method, a diversity loss $\mathcal{L}_{diversity}$ is imposed on the learned codewords, which encourages each codeword to be equally used. In contrast, in Distance-Gumbel-Softmax, we employ a larger codebook and use the rate loss $\mathcal{L}_{rate}$ to reach the target bitrate in a rate-distortion optimization sense. In this way, the real distribution of the latent features can be captured in Distance-Gumbel-Softmax.
 
 \begin{table}[!t]
\begin{center}
\caption{Evaluation on vector quantization at 3 kbps.\label{table:vq}}
\vspace{-0.2cm}
\begin{tabular}{lccc}
\toprule
 Methods & PESQ & STOI & ViSQOL\\
\midrule 
Gumbel-Softmax & 2.738 & 0.910 & 3.204 \\
Distance-Gumbel-Softmax  & \textbf{2.763} & \textbf{0.917} & \textbf{3.219}\\
\bottomrule
\end{tabular}
\end{center}
\vspace{-0.2cm}
\end{table}
 We also show the confidence score of selecting the best codeword in Fig. \ref{fig:codeword}(b), based on the learned soft probability by softmax. The Distance-Gumbel-Softmax shows obviously much higher confidence than the Gumbel-Softmax, which indicates that the learned codebook in Distance-Gumbel-Softmax has more distinct class centers. This is due to the explicit introduction of the distance map, i.e., the quantization error, into the soft probability, by which the codeword closer to the current feature is encouraged to be selected, and more distinct codewords are learned through back-propagation.
 
\begin{table}[t]
\begin{center}
\caption{Evaluation on adversarial training.\label{table:adversarial training}}
\vspace{-0.2cm}
\begin{tabular}{lcccc}
\toprule
Methods & Bitrate & PESQ & STOI  & ViSQOL\\
\midrule
TF-Codec w/o Adv. & 1 kbps & 2.085 & 0.868  & 2.742 \\
TF-Codec w Adv.& 1 kbps & \textbf{2.351} & \textbf{0.887} & \textbf{2.851}\\
\midrule
TF-Codec w/o Adv.& 3 kbps & 2.763 & 0.917  & 3.219\\
TF-Codec w Adv. & 3 kbps & \textbf{3.124} & \textbf{0.933}  & \textbf{3.510}\\
\midrule
TF-Codec w/o Adv. & 6 kbps& 3.426 & 0.949 & \textbf{3.966}\\
TF-Codec w Adv. & 6 kbps & \textbf{3.547} & \textbf{0.953}  & 3.841\\
\bottomrule
\end{tabular}
\end{center}
\vspace{-0.2cm}
\end{table}
\subsubsection{Adversarial training}
 We also conduct an ablation study to evaluate the performance of adversarial training, as presented in Table \ref{table:adversarial training}. We show the comparison with and without adversarial training at various bitrates. It is observed that with adversarial training, the PESQ, STOI and ViSQOL are largely improved, especially at 1 kbps and 3 kbps. We also observe that the feature-matching loss $\mathcal{L}_{feat}$ plays an important role in recovering high-frequency details in the generated audio in our experiment. 

\begin{figure*}[ht]
\begin{minipage}[b]{1.0\linewidth}
  \centering
  \subfigure[6 kbps]{\includegraphics[width=0.18\linewidth,height=0.14\linewidth]{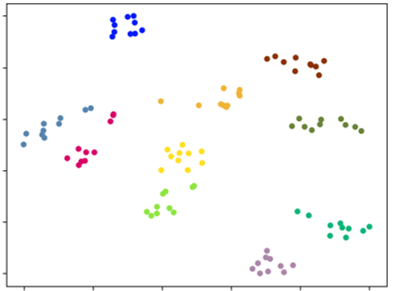}}
    \subfigure[3 kbps]{\includegraphics[width=0.18\linewidth,height=0.14\linewidth]{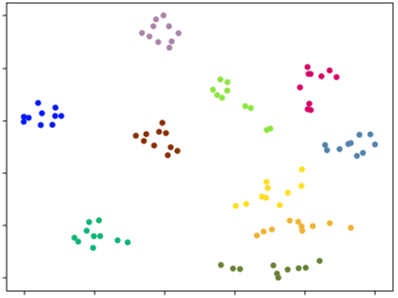}}
    \subfigure[1 kbps]{\includegraphics[width=0.18\linewidth,height=0.14\linewidth]{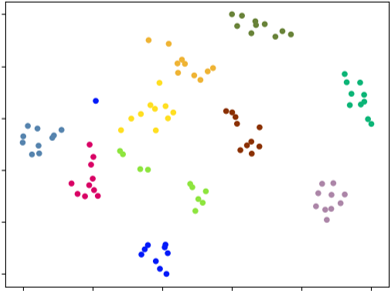}}
    \subfigure[0.512 kbps]{\includegraphics[width=0.18\linewidth,height=0.14\linewidth]{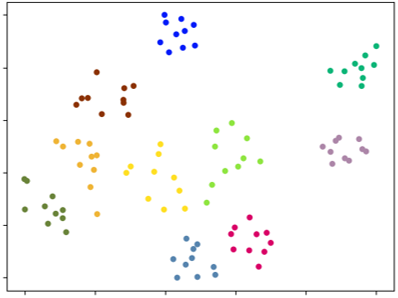}}
    \subfigure[0.256 kbps]{\includegraphics[width=0.18\linewidth,height=0.14\linewidth]{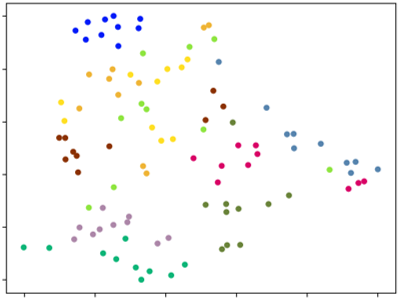}}
\end{minipage}
\caption{T-SNE of speaker information in discrete latent codes on Librispeech dataset.}
\vspace{-0.2cm}
\label{fig:vis_spk}
\end{figure*}

\begin{figure*}[ht]
\begin{minipage}[b]{0.5\linewidth}
  \centering
  \centerline{\includegraphics[width=8cm]{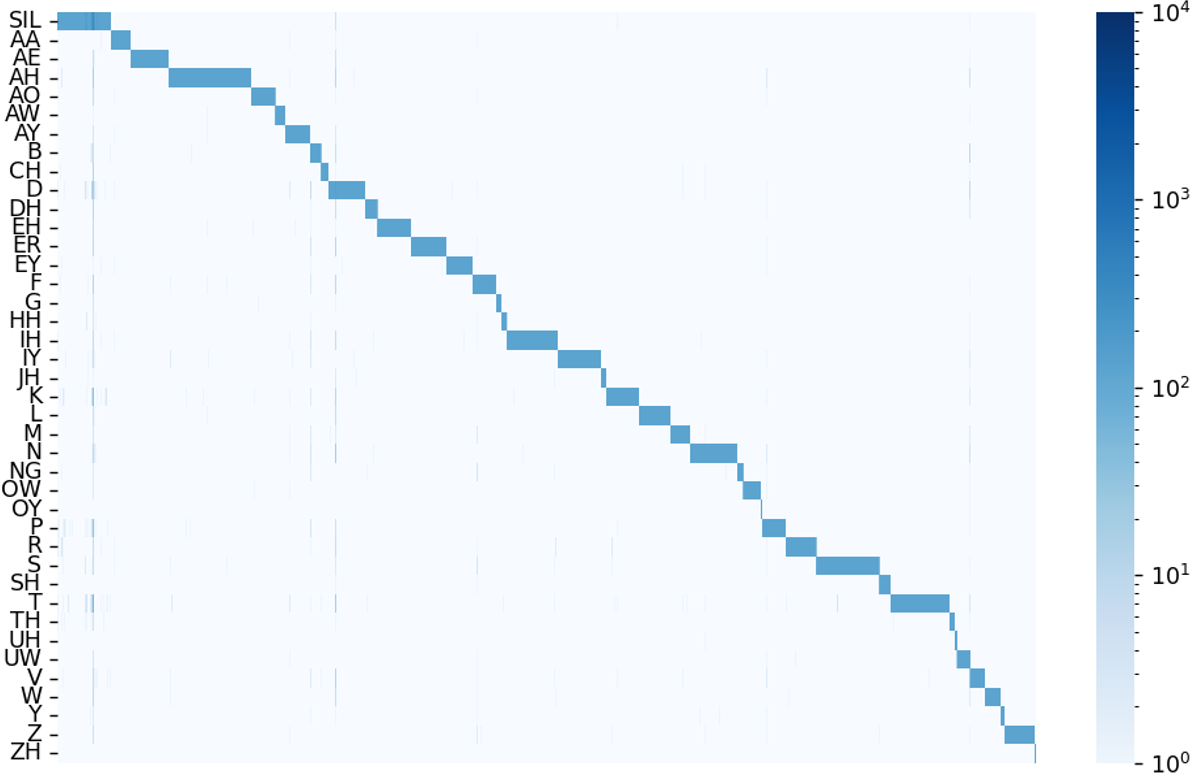}}
  \centerline{(a) 3 kbps}\medskip
\end{minipage}
\begin{minipage}[b]{0.5\linewidth}
  \centering
  \centerline{\includegraphics[width=8cm]{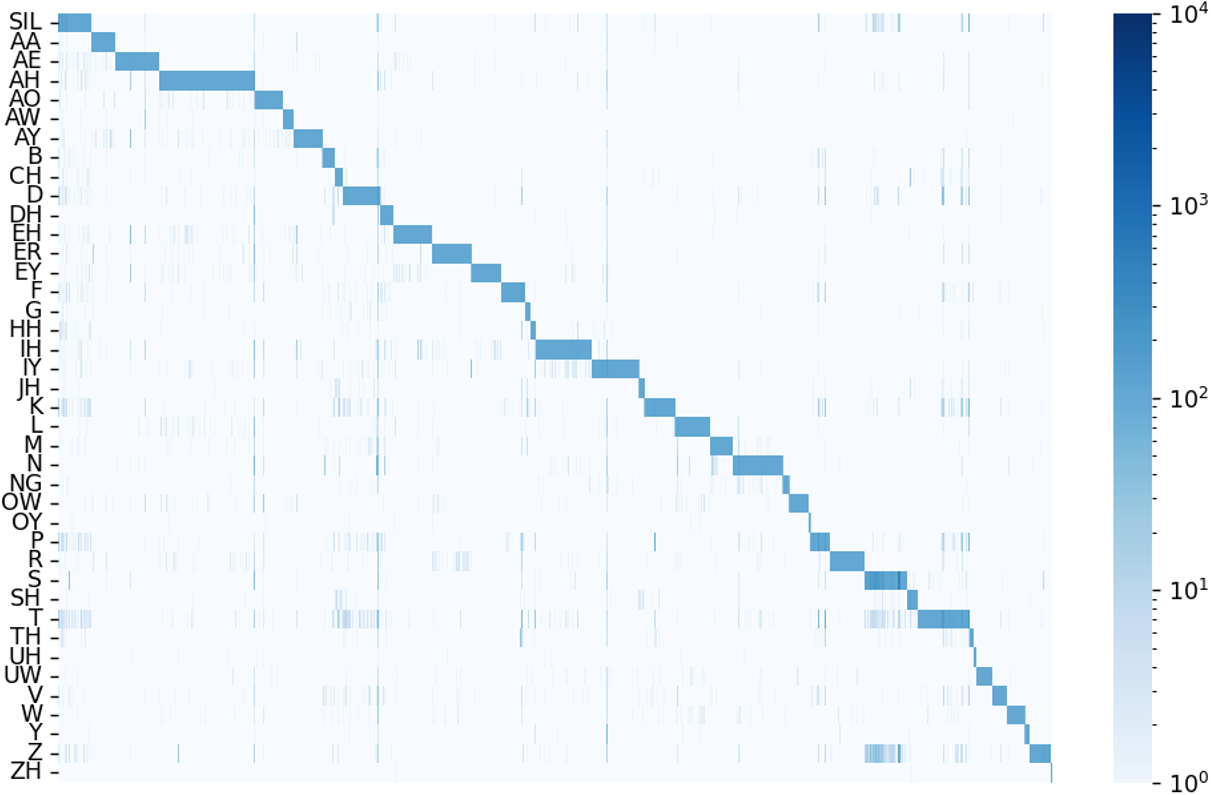}}
  \centerline{(b) 0.512 kbps}\medskip
\end{minipage}
\caption{Co-occurrence of latent codes and phonemes on LJ single speaker dataset. The horizontal axis is latent code index and the vertical axis denotes phonemes. We obtain the phoneme-level alignments with the Montreal Forced Aligner (MFA) \cite{phoneme-aligner}, using their pre-trained Librispeech acoustic model. The frame-level phoneme labels are determined by the phoneme with most occurrences in the duration of each frame.}
\vspace{-0.2cm}
\label{fig:vis_phoneme}
\end{figure*}

\subsection{Analysis}
To better understand what information is learned and encoded at various bitrates, we conduct an analysis on the learned representations and codebooks in this section. In this analysis, we disable the predictive loop and choose the discrete latent codes for visualization. 
 
 We visualize the speaker and content information contained in the learned discrete latent codes using two datasets:  (i) the Librispeech multi-speaker dataset \cite{librispeech} for speaker-related information analysis; (ii) the LJ single speaker dataset \cite{ljspeech17} for linguistic information analysis.

\textbf{Speaker information} 
We randomly select 10 speakers from Librispeech, each with 10 utterances. For each utterance, we perform a temporal average pooling on the multi-frame features, yielding a global embedding per utterance. Fig. \ref{fig:vis_spk} shows the t-SNE visualization of those speaker embeddings from 0.256 kbps to 6 kbps. We can observe that the model at high bitrates generates more compact speaker clusters, while at very low bitrates, the cluster begins to diffuse, and speakers could not be identified at 0.256 kbps. This indicates that at very low bitrates, the model turns to drop speaker-related information so as to leave bandwidth for some key information of speech. It is worth noting that the bitrate 0.256 kbps is even close to the estimated information rate of speech communication in \cite{lowbound}. At such low bitrates, linguistic information is more important than speaker variations for real-time communications.  

\textbf{Linguistic information} We evaluate the linguistic information in the discrete codes by the co-occurence map between phonemes and discrete latent codes by group vector quantization. We use all 13100 audio clips of the LJSpeech dataset spoken by the same speaker to remove the impact by speaker variations. Fig. \ref{fig:vis_phoneme} shows the co-occurence map at difference bitrates. It can be seen that these discrete latent codes, learned in a self-supervised way, are closely related to phonemes, and many latents are dedicated to specific phonemes. For example, a large number of discrete codewords are automatically allocated to specific phonemes, e.g., AH, N, S, and T. It is also observed that the distributions of the co-occurrence map for 0.512 kbps and 3 kbps are quite close to each other, indicating that the latent codes at different bitrates preserve phoneme information well. This is consistent with our hypothesis that at extremely low bitrates, the model tries to allocate the limited bandwidth to key content information (linguistic-related) in speech and drop less important information (speaker-related). 

\begin{figure}[tb]
\begin{minipage}[c]{1\linewidth}
  \centering
    \subfigure[PESQ]{\includegraphics[width=0.33\linewidth,height=0.33\linewidth]{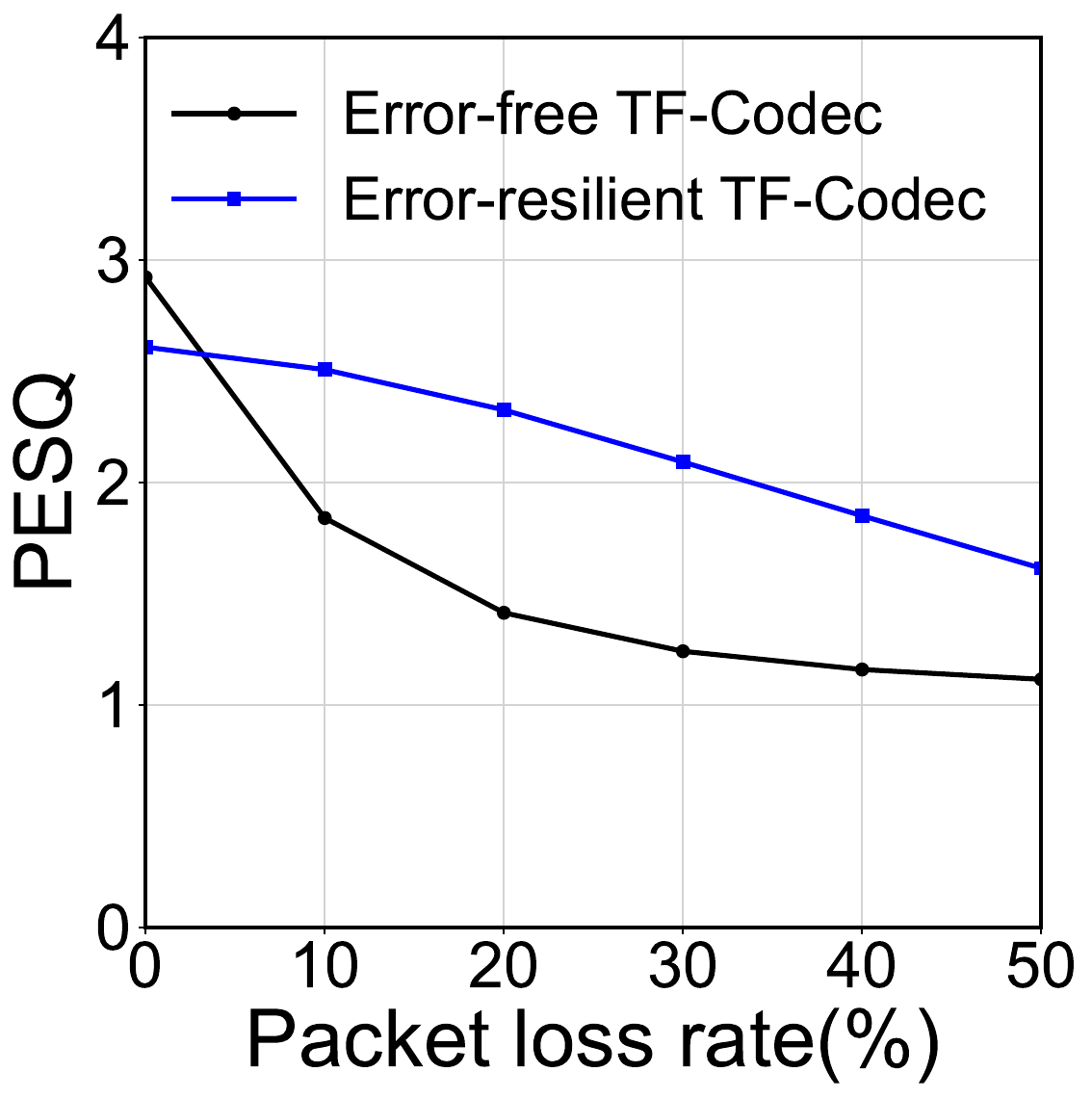}}
    \hspace{-0.2cm}
    \subfigure[STOI]{\includegraphics[width=0.33\linewidth,height=0.33\linewidth]{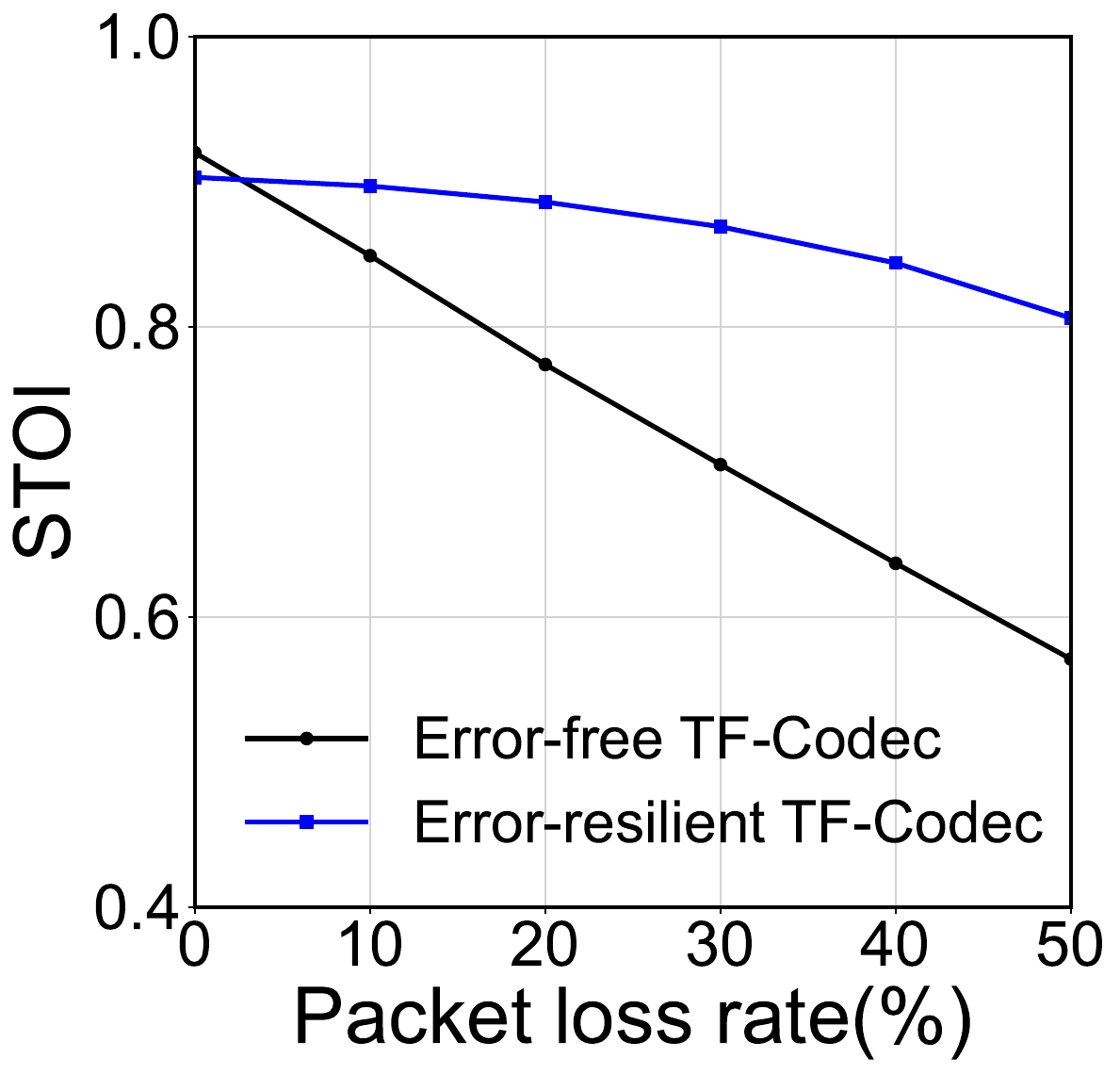}}
    \hspace{-0.2cm}
    \subfigure[PLCMOS]{\includegraphics[width=0.33\linewidth,height=0.33\linewidth]{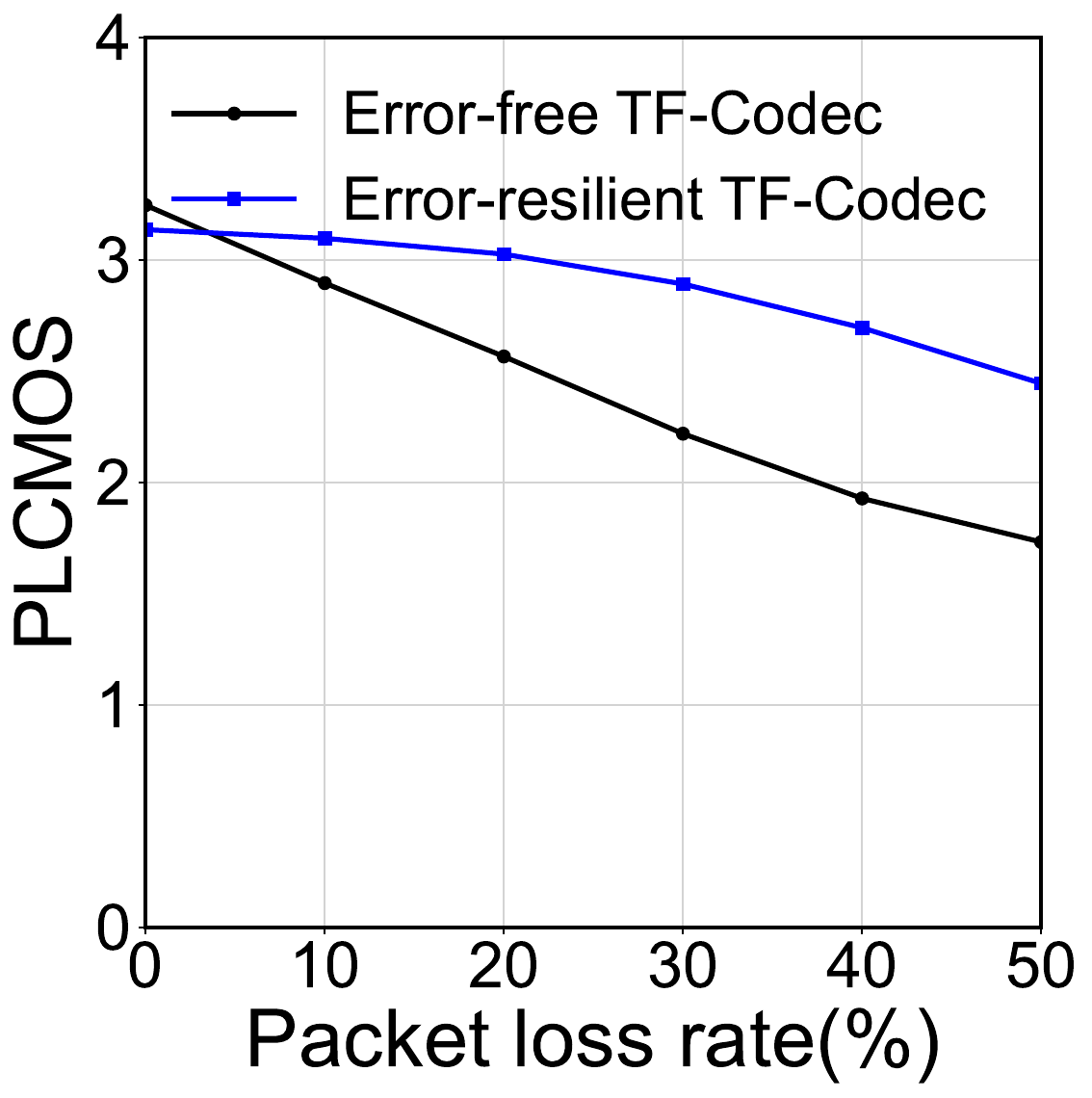}}
\end{minipage}
\caption{Evaluation of TF-Codec at packet loss rate $0\%$ to $50\%$ on synthetic test set.}
\label{fig:plc}
\end{figure}

\subsection{Robustness to Transmission Errors}
It is generally assumed that the predictive loop used to reduce temporal redundancy is sensitive to transmission errors, as it may introduce long-term error propagation. In this section, we propose some ways to improve the robustness under packet losses with loss-aware training and show some preliminary but promising results.

We simulate packet losses with a random loss rate from $\{10\%, 20\%, 30\%, 40\%, 50\%\}$, 100 hours for each category. We also simulate 390 hours of data with a WLAN packet loss pattern using three-state Markov models, similar to that in \cite{error-resilience-codec} for training. Under packet losses, the predictive loop at the encoder side operates the same as before, but in decoding, the quantized feature $\hat{\bm{x}}_t^N$ is set to zero if the packet is lost. We split each 40ms data into two packets along the channel dimension for better resilience. For evaluation metrics, besides PESQ and STOI, we also use PLCMOS\cite{plcmos}, an evaluation tool proposed in the Packet Loss Concealment Challenge at INTERSPEECH 2022 to measure the concealment quality.

In addition to training on simulated packet loss datasets, we also introduce an error-aware loss term $\mathcal{L}_{error-aware}$ on the predictor in the decoder loop, given by
\begin{equation}
\mathcal{L}_{error-aware} = \mathbb{E}_x[\frac{1}{C_{d}T}\sum_{t=1}^T||\tilde{\bm{x}}_t^P, sg(\bm{x}_t^R)||_2^2],
\label{eq:error-aware}\end{equation}
which is quite similar to $\mathcal{L}_{pred}$ in Eq. (\ref{eq:pred}) but different in that the prediction $\tilde{\bm{x}}_t^P$ is performed on the decoder-side reconstructed feature with transmission errors. This loss will force the predictor to provide an error-resilient prediction at decoding. We also find in our experiment that the residual-like feature to quantize $\bm{x}_t^N$ in predictive TF-Codec usually has low energy, and the loss of features with extremely low energy usually does not have a severe impact on the reconstruction. 

Fig. \ref{fig:plc} shows the results. It can be seen that without loss-aware training, denoted as ``Error-free TF-Codec" in Fig. \ref{fig:plc}, the quality drops sharply when there is packet loss, showing its sensitivity to transmission errors. When training with simulated packet loss, denoted as ``Error-resilience TF-Codec'', the robustness and restoration capability of TF-Codec are largely improved. However, the reconstructed audio still has perceivable artifacts, especially for burst losses, i.e., over 120ms, which can be further optimized.

It is worth mentioning that many techniques exist which can reduce the impact of transmission errors in traditional video and audio coding, such as intra-refresh in H.26X and forward error protection with redundant packets in Opus. In the future, we could consider them for further optimization towards error-resilience neural speech coding.

\subsection{Delay and Complexity}
The proposed predictive TF-Codec has  $6.37$M parameters in total, with $2.11$M for the 2D encoder, $3.44$M for the 2D decoder, and $0.82$M for the predictive loop (including predictor, extractor and synthesizer). We report the algorithm delay and real-time factor (RTF) in the following. 

\textbf{Algorithm delay} 
Algorithm delay refers to the delay caused by relying on future samples for processing the current one. All encoder and decoder layers in our codec are causal, so the algorithm delay comes from the STFT analysis window of a $40$ ms length, plus an extra $30$ ms delay as we quantize and encode four frames together.

\textbf{Real-time factor} 
RTF is calculated as the ratio between the duration of the audio and the inference time. An RTF greater than $1$ means that the system could process the data in real-time. When {running} on a single CPU (Intel$^\circledR$ Xeon$^\circledR$ Processor E5-1620 v4 3.50GHz), TF-Codec could achieve $4.1\times$ for encoding and $6.3\times$ for decoding,  achieving real-time processing.

\section{Conclusion}
We propose the TF-Codec, a low-latency neural speech codec that outperforms state-of-the-art audio codecs with very low bitrates. We introduce latent-domain predictive coding into the VQ-VAE framework to fully remove the temporal redundancy. A learnable input compression is proposed to balance the attention paid {to} main components and details in the STFT spectrum at different bitrates. We also introduce the Distance-Gumbel-Softmax {mechanism} for vector quantization, which can capture the real distribution of latent features with rate-distortion optimization. It should be noted that although speech coding is taken as an example in this paper, the proposed techniques could be extended to other audio signals such as music as well. In the future, we will make such extensions. Furthermore, we will investigate more detailed representations in terms of not only speaker and content information but also prosody and emotions. {Another interesting point to explore in the future is to make the input compression factor adaptive to input content, as different contents have different frequency responses.}


\bibliographystyle{IEEEtran}

\bibliography{mybib}

\begin{thebibliography}{10}
\providecommand{\url}[1]{#1}
\csname url@samestyle\endcsname
\providecommand{\newblock}{\relax}
\providecommand{\bibinfo}[2]{#2}
\providecommand{\BIBentrySTDinterwordspacing}{\spaceskip=0pt\relax}
\providecommand{\BIBentryALTinterwordstretchfactor}{4}
\providecommand{\BIBentryALTinterwordspacing}{\spaceskip=\fontdimen2\font plus
\BIBentryALTinterwordstretchfactor\fontdimen3\font minus
  \fontdimen4\font\relax}
\providecommand{\BIBforeignlanguage}[2]{{%
\expandafter\ifx\csname l@#1\endcsname\relax
\typeout{** WARNING: IEEEtran.bst: No hyphenation pattern has been}%
\typeout{** loaded for the language `#1'. Using the pattern for}%
\typeout{** the default language instead.}%
\else
\language=\csname l@#1\endcsname
\fi
#2}}
\providecommand{\BIBdecl}{\relax}
\BIBdecl

\bibitem{wavcodec}
W.~Kleijin, F.~Lim, A.~Luebs, and J.~Skoglund, ``Wave{N}et based low rate
  speech coding,'' in \emph{ICASSP}.\hskip 1em plus 0.5em minus 0.4em\relax
  IEEE, 2018, pp. 676--680.

\bibitem{Lyra}
W.~B. Kleijn, A.~Storus, M.~Chinen, T.~Denton, F.~S. Lim, A.~Luebs,
  J.~Skoglund, and H.~Yeh, ``Generative speech coding with predictive variance
  regularization,'' in \emph{ICASSP 2021-2021 IEEE International Conference on
  Acoustics, Speech and Signal Processing (ICASSP)}.\hskip 1em plus 0.5em minus
  0.4em\relax IEEE, 2021, pp. 6478--6482.

\bibitem{sampleRNNcodec}
J.~Klejsa, P.~Hedelin, C.~Zhou, R.~Fejgin, and L.~Villemoes, ``High-quality
  speech coding with sample {RNN},'' in \emph{ICASSP}.\hskip 1em plus 0.5em
  minus 0.4em\relax IEEE, 2019, pp. 7155--7159.

\bibitem{generative}
R.~Fejgin, J.~Klejsa, L.~Villemoes, and C.~Zhou, ``Source coding of audio
  signals with a generative model,'' in \emph{ICASSP}.\hskip 1em plus 0.5em
  minus 0.4em\relax IEEE, 2020, pp. 341--345.

\bibitem{improveopus}
J.~Skoglund and J.~Valin, ``Improving {O}pus low bit rate quality with neural
  speech synthesis,'' in \emph{Interspeech}, 2020.

\bibitem{VQ-VAE-wavenet}
C.~G\^{a}rbacea, A.~van~den Oord, Y.~Li, F.~Lim, A.~Luebs, O.~Vinyals, and
  T.~C. Walters, ``Low bit-rate speech coding with {VQ-VAE} and a {W}ave{N}et
  decoder,'' in \emph{2019 IEEE Int. Conf. Acoust Speech Signal Processing
  (ICASSP)}.\hskip 1em plus 0.5em minus 0.4em\relax IEEE, 2019, pp. 735--739.

\bibitem{disentangle}
J.~Williams, Y.~Zhao, E.~Cooper, and J.~Yamagishi, ``Learning disentangled
  phone and speaker representations in a semi-supervised {VQ-VAE} paradigm,''
  in \emph{2021 IEEE Int. Conf. Acoust Speech Signal Processing
  (ICASSP)}.\hskip 1em plus 0.5em minus 0.4em\relax IEEE, 2021.

\bibitem{soundstream}
N.~Zeghidour, A.~Luebs, A.~Omran, J.~Skoglund, and M.~Tagliasacchi,
  ``Soundstream: An end-to-end neural audio codec,'' \emph{IEEE/ACM
  Transactions on Audio, Speech, and Language Processing}, 2021.

\bibitem{cascaded}
K.~Zhen, J.~Sung, M.~Lee, S.~Beack, and M.~Kim, ``Cascaded cross-module
  residual learning towards lightweight end-to-end speech coding,'' in
  \emph{Proceedings of the Annual Conference of the International Speech and
  Communication Association (Interspeech)}, 2019.

\bibitem{scalable-codec}
K.~Zhen, J.~Sung, M.~S. Lee, and M.~Kim, ``Scalable and efficient neural speech
  coding: a hybrid design,'' \emph{IEE/ACM Transactions on audio, speech, and
  language processing}, vol.~30, pp. 12--25, 2022.

\bibitem{Configurablecodec}
T.~Jayashankar, T.~Koehler, K.~Kalgaonkar, Z.~Xiu, J.~Wu, J.~Lin, P.~Agrawal,
  and Q.~He, ``Architecture for variable bitrate neural speech codec with
  configurable computation complexity,'' in \emph{ICASSP 2022 - 2022 IEEE
  International Conference on Acoustics, Speech and Signal Processing
  (ICASSP)}, 2022, pp. 861--865.

\bibitem{TFNetCodec}
X.~Jiang, X.~Peng, C.~Zheng, H.~Xue, Y.~Zhang, and Y.~Lu, ``End-to-end neural
  speech coding for real-time communications,'' in \emph{IEEE Int. Conf. Acoust
  Speech Signal Processing (ICASSP)}, 2022.

\bibitem{VQ-VAE}
A.~v.~d. Oord, O.~Vinyals, and K.~Kavukcuoglu, ``Neural discrete representation
  learning,'' \emph{NIPS}, 2017.

\bibitem{JPEG}
G.~K. Wallace, ``The jpeg still picture compression standard,'' \emph{IEEE
  transactions on consumer electronics}, vol.~38, no.~1, pp. xviii--xxxiv,
  1992.

\bibitem{H.264/AVC}
T.~Wiegand, G.~J. Sullivan, G.~Bjontegaard, and A.~Luthra, ``Overview of the h.
  264/avc video coding standard,'' \emph{IEEE Transactions on circuits and
  systems for video technology}, vol.~13, no.~7, pp. 560--576, 2003.

\bibitem{H.265/HEVC}
G.~J. Sullivan, J.-R. Ohm, W.-J. Han, and T.~Wiegand, ``Overview of the high
  efficiency video coding ({HEVC}) standard,'' \emph{IEEE Transactions on
  Circuits and Systems for Video Technology}, vol.~22, pp. 1649--1668, 2012.

\bibitem{H.266/VVC}
B.~Bross, J.~Chen, J.-R. Ohm, G.~J. Sullivan, and Y.-K. Wang, ``Developments in
  international video coding standardization after avc, with an overview of
  versatile video coding (vvc),'' \emph{Proceedings of the IEEE}, vol. 109,
  no.~9, pp. 1463--1493, 2021.

\bibitem{DPCM}
B.~S. Atal, ``Predictive coding of speech at low bit rates,'' \emph{IEEE
  Transactions on Communications}, vol.~30, pp. 600--614, 1982.

\bibitem{ADPCM}
P.~Cummiskey, N.~S. Jayant, and J.~L. Flanagan, ``Adaptive quantization in
  differential pcm coding of speech,'' \emph{The Bell System Technical
  Journal}, vol.~52, pp. 1105--1118, 1973.

\bibitem{DVC}
G.~Lu, W.~Ouyang, D.~Xu, X.~Zhang, C.~Cai, and Z.~Gao, ``{DVC}: An end-to-end
  deep video compression framework,'' in \emph{CVPR}, 2019.

\bibitem{DeepContext}
J.~Li, B.~Li, and Y.~Lu, ``Deep contextual video compression,'' in \emph{NIPS},
  2021.

\bibitem{wavenet}
A.~van~den Oord, S.~Dieleman, H.~Zen, K.~Simonyan, O.~Vinyals, A.~Graves,
  N.~Kalchbrenner, A.~Senior, and K.~Kavukcuoglu, ``Wavenet: A generative model
  for raw audio,'' in \emph{9th ISCA Speech Synthesis Workshop}, pp. 125--125.

\bibitem{LPCNet}
V.~J.M. and S.~J., ``{LPCNet}: improving neural speech synthesis through linear
  prediction,'' in \emph{ICASSP}.\hskip 1em plus 0.5em minus 0.4em\relax IEEE,
  2019.

\bibitem{LPC}
A.~Spanias, ``Speech coding: a tutorial review,'' \emph{Proceedings of the
  IEEE}, vol.~82, no.~10, pp. 1541--1582, 1994.

\bibitem{predictivelpc}
H.~Yang, W.~Lim, and M.~Kim, ``Neural feature predictor and discriminative
  residual coding for low-bitrate speech coding,'' in \emph{ICASSP 2023 - 2023
  IEEE International Conference on Acoustics, Speech and Signal Processing
  (ICASSP)}, 2023, pp. 1--5.

\bibitem{cognitivecoding}
R.~Lotfidereshgi and P.~Gournay, ``Cognitive coding of speech,'' in
  \emph{ICASSP 2022-2022 IEEE International Conference on Acoustics, Speech and
  Signal Processing (ICASSP)}.\hskip 1em plus 0.5em minus 0.4em\relax IEEE,
  2022, pp. 7772--7776.

\bibitem{sampleRNN}
S.~Mehri, K.~Kumar, I.~Gulrajani, R.~Kumar, S.~Jain, J.~Sotelo, A.~Courville,
  and Y.~Bengio, ``Sample{RNN}: an unconditional end-to-end neural audio
  generation model,'' in \emph{ICLR}, 2017.

\bibitem{opus}
\BIBentryALTinterwordspacing
J.-M. Valin, K.~Vos, and T.~B. Terriberry, ``{Definition of the Opus Audio
  Codec},'' RFC 6716, Sep. 2012. [Online]. Available:
  \url{https://www.rfc-editor.org/info/rfc6716}
\BIBentrySTDinterwordspacing

\bibitem{celp}
M.~Schroeder and B.~Atal, ``Code-excited linear prediction (celp): High-quality
  speech at very low bit rates,'' in \emph{ICASSP'85. IEEE International
  Conference on Acoustics, Speech, and Signal Processing}, vol.~10.\hskip 1em
  plus 0.5em minus 0.4em\relax IEEE, 1985, pp. 937--940.

\bibitem{Gumbelsoftmax}
H.~Zhou, A.~Baevski, and M.~Auli, ``A comparison of discrete latent variable
  models for speech representation learning,'' in \emph{2021 IEEE Int. Conf.
  Acoust Speech Signal Processing (ICASSP)}.\hskip 1em plus 0.5em minus
  0.4em\relax IEEE, 2021.

\bibitem{vq-wav2vec}
\BIBentryALTinterwordspacing
A.~Baevski, S.~Schneider, and M.~Auli, ``vq-wav2vec: Self-supervised learning
  of discrete speech representations,'' in \emph{International Conference on
  Learning Representations}, 2020. [Online]. Available:
  \url{https://openreview.net/forum?id=rylwJxrYDS}
\BIBentrySTDinterwordspacing

\bibitem{soft-to-hard}
E.~Agustsson, F.~Mentzer, M.~Tschannen, L.~Cavigelli, R.~Timofte, L.~Benini,
  and L.~V. Gool, ``Soft-to-hard vector quantization for end-to-end learning
  compressible representations,'' \emph{NIPS}, 2017.

\bibitem{TCNN}
A.~Pandey and D.~Wang, ``T{CNN}: Temporal convolutional neural network for
  real-time speech enhancement in the time domain,'' in \emph{2019 IEEE Int.
  Conf. Acoust Speech Signal Processing (ICASSP)}.\hskip 1em plus 0.5em minus
  0.4em\relax IEEE, 2019.

\bibitem{prelu}
K.~He, X.~Zhang, S.~Ren, and J.~Sun, ``Delving deep into rectifiers: Surpassing
  human-level performance on imagenet classification,'' in \emph{Proceedings of
  the IEEE international conference on computer vision}, 2015, pp. 1026--1034.

\bibitem{attention}
A.~Vaswani, N.~Shazeer, N.~Parmar, J.~Uszkoreit, L.~Jones, A.~N. Gomez,
  {\L}.~Kaiser, and I.~Polosukhin, ``Attention is all you need,'' in
  \emph{Advances in neural information processing systems}, 2017, pp.
  5998--6008.

\bibitem{encodec}
A.~D{\'e}fossez, J.~Copet, G.~Synnaeve, and Y.~Adi, ``High fidelity neural
  audio compression,'' \emph{arXiv preprint arXiv:2210.13438}, 2022.

\bibitem{hifigan}
J.~Kong, J.~Kim, and J.~Bae, ``Hifi-{GAN}: Generative adversarial networks for
  efficient and high fidelity speech synthesis,'' \emph{Advances in Neural
  Information Processing Systems}, vol.~33, pp. 17\,022--17\,033, 2020.

\bibitem{melgan}
K.~Kumar, R.~Kumar, T.~de~Boissiere, L.~Gestin, W.~Z. Teoh, J.~Sotelo,
  A.~de~Br{\'e}bisson, Y.~Bengio, and A.~C. Courville, ``Mel{GAN}: Generative
  adversarial networks for conditional waveform synthesis,'' \emph{Advances in
  neural information processing systems}, vol.~32, 2019.

\bibitem{leaklyrelu}
B.~Xu, N.~Wang, T.~Chen, and M.~Li, ``Empirical evaluation of rectified
  activations in convolutional network,'' \emph{arXiv preprint
  arXiv:1505.00853}, 2015.

\bibitem{lsgan}
X.~Mao, Q.~Li, H.~Xie, R.~Y. Lau, Z.~Wang, and S.~Paul~Smolley, ``Least squares
  generative adversarial networks,'' in \emph{Proceedings of the IEEE
  international conference on computer vision}, 2017, pp. 2794--2802.

\bibitem{powermse}
A.~Ephrat, I.~Mosseri, O.~Lang, T.~Dekel, K.~Wilson, A.~Hassidim, W.~T.
  Freeman, and M.~Rubinstein, ``Looking to listen at the cocktail party: a
  speaker-independent audio-visual model for speech separation,'' \emph{ACM
  Transactions on Graphics (TOG)}, vol.~37, no.~4, pp. 1--11, 2018.

\bibitem{stft-consistency}
S.~Wisdom, J.~R. Hershey, K.~Wilson, J.~Thorpe, M.~Chinen, B.~Patton, and R.~A.
  Saurous, ``Differentiable consistency constraints for improved deep speech
  enhancement,'' in \emph{ICASSP}.\hskip 1em plus 0.5em minus 0.4em\relax IEEE,
  2019, pp. 900--904.

\bibitem{melloss}
A.~Gritsenko, T.~Salimans, R.~van~den Berg, J.~Snoek, and N.~Kalchbrenner, ``A
  spectral energy distance for parallel speech synthesis,'' \emph{Advances in
  Neural Information Processing Systems}, vol.~33, pp. 13\,062--13\,072, 2020.

\bibitem{evs}
M.~Dietz, M.~Multrus, V.~Eksler, V.~Malenovsky, E.~Norvell, H.~Pobloth,
  L.~Miao, Z.~Wang, L.~Laaksonen, A.~Vasilache \emph{et~al.}, ``Overview of the
  evs codec architecture,'' in \emph{2015 IEEE International Conference on
  Acoustics, Speech and Signal Processing (ICASSP)}.\hskip 1em plus 0.5em minus
  0.4em\relax IEEE, 2015, pp. 5698--5702.

\bibitem{DNSchallenge}
C.~K.~A. Reddy, H.~Dubey, V.~Gopal, R.~Cutler, S.~Braun, H.~Gamper, R.~Aichner,
  and S.~Srinivasan, ``Icassp 2021 deep noise suppression challenge,'' in
  \emph{ICASSP 2021 - 2021 IEEE International Conference on Acoustics, Speech
  and Signal Processing (ICASSP)}, 2021, pp. 6623--6627.

\bibitem{adam}
\BIBentryALTinterwordspacing
D.~P. Kingma and J.~Ba, ``Adam: {A} method for stochastic optimization,'' in
  \emph{3rd International Conference on Learning Representations, {ICLR} 2015,
  San Diego, CA, USA, May 7-9, 2015, Conference Track Proceedings}, Y.~Bengio
  and Y.~LeCun, Eds., 2015. [Online]. Available:
  \url{http://arxiv.org/abs/1412.6980}
\BIBentrySTDinterwordspacing

\bibitem{mushra}
B.~Series, ``Method for the subjective assessment of intermediate quality level
  of audio systems,'' \emph{International Telecommunication Union
  Radiocommunication Assembly}, 2014.

\bibitem{pesq}
I.~Rec, ``P.862.2: Wideband extension to recommendation p.862 for the
  assessment of wideband telephone networks and speech codecs,''
  \emph{International Telecommunication Union, CH--Geneva}, 2005.

\bibitem{stoi}
C.H.Taal, R.C.Hendriks, R.Heusdens, and J.Jensen, ``A short-time objective
  intelligibility measure for time-frequency weighted noisy speech,'' in
  \emph{ICASSP}, 2010.

\bibitem{visqol}
A.~Hines, J.~Skoglund, A.~Kokaram, and N.~Harte, ``Visqol: The virtual speech
  quality objective listener,'' in \emph{IWAENC 2012; International Workshop on
  Acoustic Signal Enhancement}.\hskip 1em plus 0.5em minus 0.4em\relax VDE,
  2012, pp. 1--4.

\bibitem{tsne}
L.~Van~der Maaten and G.~Hinton, ``Visualizing data using t-sne.''
  \emph{Journal of machine learning research}, vol.~9, no.~11, 2008.

\bibitem{librispeech}
V.~Panayotov, G.~Chen, D.~Povey, and S.~Khudanpur, ``Librispeech: an {ASR}
  corpus based on public domain audio books,'' in \emph{2015 IEEE international
  conference on acoustics, speech and signal processing (ICASSP)}.\hskip 1em
  plus 0.5em minus 0.4em\relax IEEE, 2015, pp. 5206--5210.

\bibitem{phoneme-aligner}
M.~McAuliffe, M.~Socolof, S.~Mihuc, M.~Wagner, and M.~Sonderegger, ``Montreal
  forced aligner: {T}rainable text-speech alignment using kaldi.'' in
  \emph{Interspeech}, vol. 2017, 2017, pp. 498--502.

\bibitem{ljspeech17}
K.~Ito and L.~Johnson, ``The lj speech dataset,''
  \url{https://keithito.com/LJ-Speech-Dataset/}, 2017.

\bibitem{lowbound}
S.~V. Kuyk, W.~B. Kleijn, and R.~C. Hendriks, ``On the information rate of
  speech communication,'' in \emph{ICASSP}, 2017, pp. 5625--5629.

\bibitem{error-resilience-codec}
H.~Xue, X.~Peng, X.~Jiang, and Y.~Lu, ``{Towards Error-Resilient Neural Speech
  Coding},'' in \emph{Proc. Interspeech 2022}, 2022, pp. 4217--4221.

\bibitem{plcmos}
L.~Diener, S.~Sootla, S.~Branets, A.~Saabas, R.~Aichner, and R.~Cutler,
  ``{INTERSPEECH 2022 Audio Deep Packet Loss Concealment Challenge},'' in
  \emph{Proc. Interspeech 2022}, 2022, pp. 580--584.

\end{thebibliography}













\vfill

\end{document}